\documentclass[twocolumn]{revtex4}
\usepackage[colorlinks,bookmarksopen,bookmarksnumbered,citecolor=magenta,urlcolor=red]{hyperref}
\usepackage{graphicx}
\usepackage{amsmath}
\usepackage{natbib}
\usepackage{subfigure}
\usepackage{xcolor}
\usepackage[percent]{overpic}
\usepackage{tikz}
\usepackage{amsmath}
\usepackage{url}
\usepackage{times}
\usepackage{epstopdf}
\begin{document}
\bibliographystyle{abbrvnat}
	
\title{Eddy-freshwater Interaction using Regional Ocean Modeling System in the Bay of Bengal}
	
	
	

\author{Nihar Paul$^{1}$, Jai Sukhatme$^{1,2}$, Bishakhdatta Gayen$^{1,3}$, and Debasis Sengupta$^{1,2}$}

\affiliation{$^1$Centre for Atmospheric and Oceanic Sciences, Indian Institute of Science, Bengaluru - 560012, India,}
\affiliation{$^2$Divecha Centre for Climate Change, Indian Institute of Science, Bengaluru - 560012, India,}
\affiliation{$^3$Mechanical Engineering Department, University of Melbourne, Australia.}		
	
	
\begin{abstract}
{
\noindent Eddy-freshwater interaction is studied in the north Bay of Bengal (BoB) with a high-resolution simulation using the Regional Ocean Modeling System. Following observations, the model simulates the trapping and homogenization of river water by a cyclonic mesoscale eddy on a sub-monthly time scale from October-November of 2015. As fresh river water is trapped in the eddy, it is characterized by strong vertical and lateral gradients in salinity. Within a few weeks, these gradients relax along with the progressive homogenization of freshwater within the eddy. A mixed layer salinity budget shows the importance of ageostrophic vertical advection in addition to lateral advection during the evolution of salinity within the eddy. An analysis of the eddy kinetic energy (EKE) budget in the upper ocean indicates the development of barotropic and baroclinic instabilities. The vertical profiles of EKE conversion terms reveal that the surface freshwater was involved in the evolution of baroclinic instability within the mixed layer. In addition, an eddy available potential energy (EPE) budget shows that the entrainment of the river water raises the EPE, which is due to an increase in lateral salinity gradients across the eddy during the trapping event. Subsequently, the salinity homogenization leads to a decrease in the EPE, and its rate of decay is modulated by a correlation between surface buoyancy fluxes and density anomalies. Finally, reanalysis data show similar trapping and homogenization events across multiple years, highlighting the importance of this mechanism of subseasonal freshwater evolution in the BoB.
}
\end{abstract}
	
\vskip 0.25 truecm

\maketitle
\section*{Significance Statement}
{\noindent The Bay of Bengal (BoB) receives a large amount of freshwater from rivers during the summer monsoon season. This freshwater forms a very shallow layer on the surface of the Bay, and the shallow salinity-dominated stratification is important for regional air-sea interaction. However, accurate simulation of the space-time evolution of upper ocean salinity in the BoB remains a challenge for numerical models. In this work, we elucidate one particular mechanism that involves the trapping of freshwater by a cyclonic eddy and its homogenization on a sub-monthly timescale along the western boundary of the north BoB through a high-resolution numerical simulation. The mesoscale dynamical processes involved in homogenization are identified and, for the first time, allow for an understanding of surface salinity evolution on a subseasonal scale.}

\section{Introduction}
\label{intro}
\noindent The Bay of Bengal (BoB) is a semi-enclosed marginal sea ($6^\circ-22^\circ$E, $80^\circ-100^\circ$N) located in the northeastern tropical Indian Ocean; it is the largest and freshest Bay in the tropical ocean, with sea-surface salinity values between $24-32$ psu during late summer and autumn \citep{rao2003seasonal,sengupta2016near,sree2020quasi}. The major sources of freshwater to northern BoB are Ganga-Brahmaputra-Meghna (GBM), Irrawaddy, Godavari, Mahanadi, and Krishna rivers. The mean annual climatological discharge peaks during the monsoonal season (June-September), the largest being for GBM and Irrawaddy rivers with rates of $\sim 8.7\times 10^4$ m$^3$ s$^{-1}$ and $3.4\times10^4$ m$^3$ s$^{-1}$, respectively \citep{dai2002estimates,papa2010satellite}. The river water from GBM and Irrawaddy forms a tenacious, very shallow, highly stratified surface layer in north BoB \citep{sengupta2016near}, and this low salinity surface water has a profound influence on air-sea interactions and regional monsoonal rainfall \citep{gadgil2003indian,shenoi2002differences,goswami2016monsoons}.	In addition, this river water in the north BoB also controls the mixed layer depth (MLD) and allows the formation of a barrier layer \citep{sengupta2001oscillations,vinayachandran2002observations,prend2019impact} defined by the region between the MLD and the deeper isothermal layer during late summer and autumn. Further, the strong stratification and sea-surface temperature (SST) cooling due to surface heat loss in winter enable the formation of multiple fine-scale (less than 10 m) inversion layers in temperature from November to February \citep{thad,thadathil2016surface,shroyer2016modification}. Thus, the convectively stable stratified warm sub-surface layer acts as a reservoir of heat which affects the SST \citep{mahadevan2016freshwater} and can reduce the maximum depth of storm-induced vertical mixing with little or no sea-surface temperature cooling, thereby influencing the intensification of post-monsoon tropical cyclones in the BoB \citep{sengupta2008cyclone,balaguru2014increase,neetu2019premonsoon,chaudhuri2019response}.\\

\noindent Mesoscale eddies are ubiquitous in the BoB, and instabilities facilitate energy conversion from the seasonally varying mean flow to eddies in the western boundary region of the Bay \citep{chen2012features,chen2018origins}. Mesoscale eddies play a crucial role in the transport of heat, salt, and passive scalars \citep{sree2018subseasonal,busireddy2018observational,paul2020seasonality,paul2021eddy}. For example, \cite{sree2020quasi} showed during the quiescent (dry and calm) phase of the Indian summer monsoon, river water is carried from the northern BoB towards the open ocean by the geostrophic flow associated with mesoscale eddies. BoB eddies influence upper ocean thermodynamics, biological productivity, and the oxygen minimum zone (OMZ) \citep{prasanna2002bay,sarma2019role,thushara2019vertical,d2020structure}. Given their importance, the statistical properties and vertical structure of eddies in the Bay have been extensively studied using altimetry and Argo floats \citep{chen2012features,cui2016statistical}. \\

\noindent Even though many aspects of eddies in the BoB have been studied, their interaction with freshwater and the dynamical processes involved have not been explored in detail. In particular, oceanic flow is dominated by the geostrophic eddy ﬁeld, and the interaction with winds, buoyancy fluxes (surface freshwater salt flux, and heat flux) influence the various reservoirs and sinks of eddy kinetic energy \citep{ferrari2009ocean,zippel2022parsing}, and eddy available potential energy \citep{storch2012estimate,bishop2020global}. In the context of BoB sea-surface salinity, freshwater from rivers and monsoon rain forms a shallow surface layer that has very low salinity along the northern rim of the Bay and in the open ocean. River water is stirred into the interior of the north BoB by the surface flow, which is characterized by energetic mesoscale eddies. Surface salinity in the open ocean is generally lowest in August-October, but increases by December. Previous studies using eddy-permitting models suggest that horizontal advection within the mixed layer conveys freshwater out of the north Bay, but vertical diffusive processes lead to ``homogenization'' or an increase in salinity on the timescale of a few months \citep{akhil2014modeling,benshila2014upper}. Similar modeling studies using the Hybrid Coordinate Ocean Model and the Navy Coupled Ocean Data Assimilation also suggest that vertical salt fluxes are primarily responsible for counterbalancing near-surface freshening \citep{wilson2016assessment}. It is to be noted most of these studies focused on a basin-scale and seasonal timescale, showcasing the dominating role of vertical diffusion. Studies by \cite{Amala} have shown that the Lagrangian change in the salinity from August to October of 2013 can increase the parcels' saltiness in this relatively short period with southward advection. Using a host of {\it in situ}, satellite and reanalysis data, \cite{paul2021eddy} showed how a mesoscale eddy could trap freshwater and homogenize it within a time scale of a month along the western boundary of north BoB during October-November of 2015. They highlighted that horizontal chaotic mixing is one of the mechanisms by which freshwater (within the eddy) can mix with the salty water (outside the eddy) at the surface and that the mixed layer salinity evolves qualitatively in agreement with horizontal advection.\\ 

\noindent In this work we aim to elucidate the dynamical processes that are active during eddy-freshwater interaction, akin to the adjustment of a freshwater eddy in the salty environment of the BoB. We begin by describing the model configuration used to numerically simulate the trapping and homogenization of freshwater in Section \ref{mod-conf}. We then characterize the eddy-freshwater interaction from the model data along with its shortcomings in Section \ref{eddy-fresh-int}. The repeatability of eddy-freshwater trapping and homogenization events is presented across multiple years from 2011 to 2019 from reanalysis data. In Section \ref{mld-bud}, we study the evolution of mixed-layer salinity during the trapping and homogenization of freshwater. In Section \ref{sources-sink}, we move to energy sources and sinks, essentially quantifying the energy conversion for an Eddy Kinetic Energy budget. The budget for an Eddy Available Potential Energy is then presented in Section \ref{epe-bud}. Finally, we end the manuscript with a conclusion and summary of the results.   

\section{Model configuration}
\label{mod-conf}

\noindent We use Regional Ocean Modeling System (ROMS) to study the eddy-freshwater interaction in the north BoB. This is a free surface, terrain following, hydrostatic primitive equation model with Boussinesq approximation. The model solves the primitive equations in an orthogonal curvilinear coordinate system using a split-explicit time-stepping scheme \citep{haidvogel2000model,shchepetkin2005regional} on the Arakawa C grid. The model domain extends from 79$^\circ$E-99$^\circ$E and 10$^\circ$N-24$^\circ$N with horizontal resolution of $\frac{1}{36}^\circ$ ($\sim$ 3 km) and 50 vertical sigma levels in the stretched s-coordinate system shown in Figure S1. The vertical stretching parameters near-surface ($\theta_s$) and bottom ($\theta_b$) are 9 and 2.5, respectively, and have been chosen to accommodate about $23-24$ levels in the top 50 m of the water column to get a high-resolution vertical structure of the eddy. The model has high-resolution bathymetry and coastline information \citep{smith1997global}. Harmonic and bi-harmonic mixing schemes are used for the horizontal mixing of momentum and tracers along the geopotential surfaces \citep{haidvogel1999numerical}, respectively. The model uses the K-profile parameterization mixing scheme \citep{large1994oceanic} for vertical mixing. The model domain has one open boundary on the southern edge at 10$^\circ$N and three closed boundaries on the remaining edges. The no-slip condition has been used for the flow velocities at the closed boundaries, and the tracer fluxes are set to zero, normal to the wall and the bottom of the ocean basin. The open boundary uses Chapman condition \citep{chapman1985numerical} for surface elevation; Flather condition \citep{flather1976tidal} for barotropic velocities; mixed-radiation nudging condition for baroclinic velocities and tracers \citep{orlanski1976simple,marchesiello2001open,di2003seasonal} with stronger nudging on inflow (time scale of 1 day) than on outflow (10 days), and a gradient boundary condition for mixing turbulent kinetic energy. Quadratic drag with a coefficient of $10^{-3}$ is employed for momentum dissipation at the bottom of the domain. \\ 

\noindent The model is initialized with Nucleus for European Modeling of the Ocean (NEMO) reanalysis \citep{ferry2010mercator,ferry2012nemo,lellouche2018recent,chassignet2018new} at a resolution of $\frac{1}{12}^\circ$ ($\sim$8 km) interpolated to a resolution of $\frac{1}{36}^\circ$ ($\sim$3 km) for 1st October of 2015. Further, daily NEMO reanalysis data was provided at the open boundary conditions. We used COARE 3.0 bulk-flux algorithm to compute surface momentum fluxes and latent and sensible heat fluxes \citep{fairall1996cool,fairall2003bulk}. We provided zonal and meridional winds at 10 m (6-hourly), sea-level pressure (6-hourly) from ERA-interim \citep{dee2011era}; incident shortwave radiation, net longwave radiation, air temperature, and specific humidity at 2 m from daily TropFlux surface fluxes \citep{kumar2012tropflux,kumar2013tropflux}; and three hourly rainfall from the multi-satellite Tropical Rainfall Measuring Mission (TRMM) 3B42v7 data \citep{huffman2007trmm} for surface boundary forcing. The tributaries are marked as a source point along the continental shelf at the mouth of all the major rivers connecting north Bay. The river water discharges for the GBM and the Irrawaddy rivers are obtained from the altimeter-based monthly estimates of \cite{papa2012ganga} for 2015. The discharge from other rivers (Krishna, Cauvery, Mahanadi, Godavari, Subernarekha, Brahmani, and Ponnaiyar) is obtained from long-term monthly climatology estimates \citep{dai2002estimates,dai2017dai}. The total discharge for each river is distributed equally amongst individual source points shown in the legend of Figure S1. The discharge is distributed equally in the first ten sigma levels at the source points (2-5 m). The temperature and salinity at the source points are specified from the monthly climatology of World Ocean Atlas-2018 (WOA-18) \citep{garcia2019world} and Soil Moisture Active Passive (SMAP) sea-surface salinity (2015-2020) \citep{fore2016combined}. The simulation has been run for 91 days till December 30, i.e., for a season October-December, thus allowing for the interaction of cyclonic eddy with freshwater in the Bay.

\section{Eddy-freshwater interaction}
\label{eddy-fresh-int}

\noindent We analyze the output data of the model simulation to study the interaction of cyclonic eddy and the freshwater on the west coast of BoB during October-November for the year 2015 as described by {\it in situ}, satellite and reanalysis in \cite{paul2021eddy}. Concerning the particular eddy-freshwater interaction event, the evolution of sea-surface salinity in the ROMS simulation is shown in Figure \ref{f1}. Before proceeding forward, a comparison between seasonal surface currents, sea-surface temperature, and sea-surface salinity for the observation, NEMO reanalysis, and model data averaged from October to December is shown in Figure S2. The seasonal surface currents from Ocean Surface Currents Analyses Real-time (OSCAR)  \citep{bonjean2002diagnostic} are captured well in a qualitative sense, as seen in Figure S2(a)-(c). Sea-surface temperature (SST) from reanalysis and the model data is compared with the Group for High-Resolution Sea Surface Temperature (GHRSST) in Figure S2(d)-(f); here, we notice that while the pattern of SST is similar in all products, both reanalysis and the model tend to be cooler than the observed GHRSST values, especially in the north BoB. Similarly, Figure S2(g)-(i) shows a comparison of sea-surface salinity (SSS) from satellite data, reanalysis, and the model data. Missing satellite data near the coasts makes comparison difficult, though the model and reanalysis appear fresher than the SMAP SSS values. But, again, the seasonal mean of sea-surface salinity is captured well by the model. Thus, in all, the mean seasonal state of the Bay is qualitatively reproduced by the ROMS run, which lends confidence to assessing the subseasonal phenomenon of trapping and homogenization in this modeling framework. In the model simulation; freshwater enters into the eddy, is stirred, it does not form lobes (except on specific dates, Figure \ref{f1}(a),(b)) around the eddy, but the intrusion of salty water from different corners is conspicuous; for example, on 26/10, salty water is found to be intruding from the southern edge of the eddy. A similar observation is noticed on 1/11, where saline water enters from the northeast corner and further detaches from the boundary current on 13/11. However, in the reanalysis data shown in Figure S3, freshwater enters into the eddy, is stirred, and wrapped in the mesoscale eddy from all corners surrounded by salty water, forming lobe-like structures and detaching from the East India Coastal Current (EICC) that carries river water equatorward along the western boundary of the Bay. In essence, the model qualitatively matches with reanalysis data, although the actual values of SSS differ slightly. Indeed, it is to be kept in mind NEMO reanalysis is a data assimilated run with negative biases concerning SMAP SSS in the BoB. \\

\noindent The vertical structure of the eddy with profiles of meridional velocity and TS ($\theta$, salinity) averaged over 28/10-21/11 are shown in Figure \ref{f1}(g), (h). The eddy is confined to the upper 150 m with $|v|_{max}\sim$ 35 cm s$^{-1}$ with inversion in temperature at a depth of $10-15$ m ($\sim 1^\circ$C) and trapping of salinity depth of $5-15$ m ($<30$ psu). The characteristics of the model eddy are in fair agreement with reanalysis \cite[Figure 6 of][]{paul2021eddy}, with the mean maximum velocity, temperature inversion depth, and salinity-trapping depth with 35 cm s$^{-1}$, $\sim$ $10-15$ m, and $15-20$ m ($<29$ psu), respectively. The evolution of salinity in the eddy is demonstrated in Figure \ref{f2}(a)-(c); these maps are averages over three days that span the homogenization process, i.e., from October to November 2015. As is evident, the overall salinity increases over this period of a month. The homogenization of surface freshwater is captured more quantitatively in Figure \ref{f2}(d) via the probability density function (pdf) of surface salinity for early (20/10-22/10), middle (4/11-6/11), and late (19/11-21/11) stages within the eddy taken over a square box of 300 km $\times $ 300 km. Initially, as seen in the blue curve, the pdf has a peak at low salinity (around 25 psu) indicative of the trapped freshwater. As the homogenization progresses, this peak shifts rightward and the pdf becomes taller and narrower in character. Note that the secondary peaks (above 30 psu) on 19/11-21/11 in the pdf are due to the intrusion of salty water into the box. The mean salinity increases and its variance gradually decreases with time from early to the late stages from $\sim$27 to 30 psu, respectively --- i.e., they indicate a subseasonal homogenization of SSS. A section extending from 84$^\circ$E$-89^\circ$E passing through the center of the eddy and the vertical profile of Rossby number (defined by the ratio of relative vorticity to Coriolis parameter --- $\frac{\zeta}{f}$), are shown in Figure S4 for 20/10 (early), 7/11 (middle) and 21/11 (late) stages of homogenization. The cyclonic eddy is characterized by positive $\frac{\zeta}{f}$ in the range of $0.1-0.5$ enhanced at the edges with small secondary vortex structures around the center, as evident in Figure S4(b). Vertically, it is characterized by thin fine-scale enhanced $\frac{\zeta}{f}$ filamentous structure (with positive and negative signs) on the eddy edges extending to $40-50$ m as shown in Figure S4(e), (f). \\

\noindent The evolution of potential temperature ($\theta$), salinity (S), squared Brunt-v\"{a}is\"{a}l\"{a} ($N^2=-\frac{g}{\rho_0}\frac{\partial \sigma_\theta}{\partial z}$), and vertical shear ($S^2=(\frac{\partial u}{\partial z})^2+(\frac{\partial v}{\partial z})^2$) are shown in Figure \ref{f3}(a)-(c), (d)-(f), (g)-(i), and (j)-(l). The eddy is characterized by a thermocline with inversion in the potential temperature which develops in time from 20/10 (with minimal inversion; Figure \ref{f3}(a)), 26/10-21/11 (pronounced inversion; Figure \ref{f3}(b),(c)) confined to upper $10-15$ m. The subsurface water is warmer by $\sim$0.5$^\circ$C to 2$^\circ$C on 7/11 and 21/11, respectively. Figure \ref{f3}(d)-(f) shows the evolution of salinity in the eddy on the dates mentioned earlier. Initially, two thin lenses of freshwater ($\le28$ psu shown for the depth of 30 m) are confined to the upper $5-8$ m; these spiral inward on 20/10 and gradually get stirred and subsequently mixed inside the eddy. Interestingly, the near-surface salinity below 32 psu contour on 20/10 horizontally aligns as the eddy stirs it, merges on 7/11, and with the progress of time, homogenizes to salinity 29-30 psu on 21/11. As the freshwater interacts with eddy the stratification which was initially very high (N$^2_{max}=1.4\times10^{-2}$ s$^{-2}$ on 20/10) gradually reduces in time from 7/11 to 21/11 (N$^2_{max}=3.2\times10^{-3}$ s$^{-2}$---1.1$\times10^{-3}$ s$^{-2}$) as seen in Figure \ref{f3}(g)-(i). The vertical shear, shown in Figure \ref{f3}(j)-(l), is characterized by low values (S$^2_{max}=2\times10^{-3}$ s$^{-2}$) on 20/10, gradually increases by 7/11 (S$^2_{max}=3.1\times10^{-3}$ s$^{-2}$) and then is reduced by 21/11 (S$^2_{max}=5.71\times10^{-4}$ s$^{-2}$). The mixed layer depth (MLD) is defined as the depth where $\sigma_\theta$ exceeds its surface value by 0.125 kg m$^{-3}$ \citep{kara2000optimal} is shown in all the sub-panels of Figure \ref{f3}. With trapping and subsequent homogenization, the MLD below the freshwater lenses within the eddy increases from 2-3 m (20/10) to 15-20 m (21/11). These features captured in the ROMS simulation follow the pathway delineated by observations in \cite{paul2021eddy}. Of course, with all the fields available in the model, we are now in a position to assess energy budgets and examine the processes that dominate the dynamics of eddy-freshwater interaction, or in other words, how the adjustment of freshwater eddy progresses over time.

\subsection{Examples of trapping and homogenization in other years}

\noindent Before performing the energy budget analysis, we first establish the robustness of the homogenization process described here. We identify multiple such events using NEMO reanalysis in different years from 2011 to 2019. These examples from 2011, 2012, and 2017 point to the repeated occurrence of trapping and homogenization events during the post-monsoon season when freshwater is discharged from the river mouth of GBM and transported along the western boundary of the BoB. To outline the trapping and homogenization events across different years, we present the evolution of the probability density function of sea-surface salinity of the eddy 
in Figure \ref{f4}. Starting in November of 2011, Figure S5(a) shows the interaction of freshwater with an anti-cyclonic eddy of size $\sim$ 200 km, where trapping and homogenization occur over 25 days. The freshwater is cooler than the ambient salty water in the Bay, as depicted in the snapshots of Figure S5(b). The meridional velocity and vertical profiles of potential temperature ($\theta$) and salinity averaged over the trapping period are shown in Figures S6(a) and (b), respectively. The vertical structure of the eddy has a westward tilt, and its signature persists to a depth of 160 m (where $|v|>10$ cm s$^{-1}$). The freshwater ($<28$ psu) forms a very thin layer at the top of the eddy, of the order $5-10$ m, and $\theta$ shows inversion in temperature of $\sim$1.5$^\circ$C between $10-15$ m. The second example is from October 25 to November 19, 2012. In this case, the evolution of low salinity water and $\theta$ is shown in Figure S7(a) and (b), respectively. The trapping and homogenization process happened over a month, and cold water is seen to be advected in late October onward across the western boundary into the eddy. Once again, the eddy signature persists to a depth of $\sim$100 m (where $|v|>10$ cm s$^{-1}$) as shown in Figure S8(a). The inversion in temperature ($\sim$0.5$^\circ$C) and trapping depth of low salinity water ($<29$ psu) are $\sim 10-15$ m and 15 m, respectively. The third example is from 2017 when freshwater gets trapped into a cyclonic eddy of $\sim$300 km in diameter, as shown in Figure S9(a). Further, cold water advected in late October from the western boundary into the eddy and formed a cool surface layer on the top of the eddy, as shown in Figure S9(b). The eddy has a strong signature and persists to a larger depth of 300 m (where $|v|<20$ cm s$^{-1}$), as seen in Figure S10(a). The inversion in temperature-depth ($\sim$0.5$^\circ$C) and freshwater depth ($<$30 psu) for the eddy vary $\sim 10-15$ m and $5-15$ m, respectively shown in Figure S10(b). The width of the pdf of these examples decreases, shifts right and becomes taller with homogenization except for 2017, shown in Figure \ref{f4}(a), (b). Summarizing these examples, the mean saltiness ($\mu$) increases, and the standard deviation ($\sigma$) decreases with mixing between eddy-trapped freshwater and ambient seawater shown in Figure \ref{f4}, thus clearly demonstrating the homogenization of SSS and its repeatable occurrence on a subseasonal timescale. In the next section, we move to the evolution of mixed layer salinity by these processes.      

\section{Mixed layer salinity budget}
\label{mld-bud}
\noindent The mixed layer salinity budget is developed based on the previous work by \cite{feng1998upper,yuhong2013impact} as follows,
\begin{equation}
\begin{split}
\underbrace{\frac{\partial S}{\partial t}}_{Tendency}=\underbrace{-(u\frac{\partial S}{\partial x}+v\frac{\partial S}{\partial y})}_{Horizontal\: advection}+\underbrace{\frac{S_0(E-P)}{h}}_{surface\: freshwater\: salt \:flux}\\+\underbrace{residue}_{Vertical\: entrainment + mixing}.
\end{split}
\label{eqn1}
\end{equation}

 \noindent Here, $S$, $S_0$ are the mixed layer and sea-surface salinity at $z=-0.6$ m, $u$ and $v$ are the mixed layer's zonal and meridional velocities, $E-P$ represents the evaporation minus precipitation rate (m s$^{-1}$), and $h$ (m) is the mixed layer depth. The term $residue$ denotes the contribution from vertical entrainment (including vertical advection) and mixing/diffusion. In Figure \ref{f5}(a)-(f), we show the difference between the mixed layer salinity tendency term and surface freshwater salt flux to showcase the contribution of horizontal advection as well as vertical entrainment and mixing/diffusion term, whereas Figure \ref{f5}(g)-(l) shows the horizontal advection of mixed layer salinity. The difference of mixed layer salinity tendency term and the surface freshwater salt flux is in accord with the horizontal advection on 20/10, 26/10 (Figure \ref{f5}(a),(g) \& \ref{f5}(b),(h)) but quite starkly, on 1/11, 7/11, 13/11, and 21/11, there is a incongruity between these terms (Figure \ref{f5}(c),(i), \ref{f5}(d),(j), \ref{f5}(e),(k), \& \ref{f5}(f),(i)) around the eddy. In essence, while horizontal advection is important and dominant in the early days, the $residue$ also plays a significant role in the middle and late stages during the homogenization of SSS. Interestingly, alternating bands of negative (blue) and positive (red) tendency are also observed across the edges of the eddy on 20/10, 26/10, and 7/11 in both the upper and lower corresponding panels of Figure \ref{f5}. In contrast, these patterns are present throughout in horizontal salinity advection in Figure \ref{f5}(g)-(k). These patterns suggest the development of wave-like processes or instabilities, and their consequences will be examined in the next section.   

\section{Energy sources and sinks}
\label{sources-sink}	
\noindent We now examine the eddy-mean flow interaction via energy exchanges between different reservoirs during trapping and homogenization. In a closed oceanic basin, four conversions take place as a part of the Lorenz Energy Cycle \citep[LEC;][]{lorenz1955available,hughes2009available} as discussed in the schematic of Figure 8 of \cite{boning1992eddy}. They are denoted by T1, T2, T3, and T4. T1 represents the conversion of mean kinetic to mean potential energy (KEM to PEM), T2 is the conversion of mean to eddy potential energy (PEM to EPE), T3 the conversion from eddy potential to eddy kinetic energy (EPE to EKE), and T4 is the work of the Reynolds stresses against the mean shear (KEM to EKE). Following \citet{harrison1978energy,hall1986diagnostic}, we show the conversion terms integrated within a volume in the eddy and present their contributions during SSS homogenization. Note that we have also accounted for the effect of wind and buoyancy while deriving these estimates. 

\subsection{Eddy Kinetic Energy}

\noindent We begin by analyzing eddy kinetic energy where the variables $\mathbf{u}=(\mathbf{u}_h,w)=(u,v,w)$ are decomposed into a background mean flow and eddy perturbation component. The mean flow ($\mathbf{\bar{u}}$) is the average over a season from October to December, the overbar denotes the temporal mean. The perturbation components ($\mathbf{u'}$) are the deviation of the daily velocity fields from the mean flow. The eddy kinetic energy (EKE) is defined by,
\begin{align}\label{eqn2}
&k_e=\rho_0\frac{\overline{\mathbf{u}^{\prime 2}}}{2}\equiv \rho_0\left(\frac{\overline{u^{\prime 2}+v^{\prime 2}}}{2}\right).
\end{align}

\noindent Here, $\rho_0=1025$ kg m$^{-3}$. The evolution of EKE ($k_e$) reads \citep{storch2012estimate},

\begin{equation}\label{eqn3}
\begin{split}
\frac{\partial k_e}{\partial t}+\boldsymbol{\nabla}\cdot(k_e\boldsymbol{\bar{u}})+\boldsymbol{\nabla}\cdot\left[ \frac{\rho_0}{2}\overline{\boldsymbol{u^\prime}(u^{\prime2}+v^{\prime2})}\right]+\boldsymbol{\nabla}\cdot\overline{p^\prime\boldsymbol{u^\prime}}\\=-\rho_0\overline{u^\prime\boldsymbol{u}^\prime}\cdot \boldsymbol{\nabla}\bar{u}-\rho_0\overline{v^\prime\boldsymbol{u}^\prime}\cdot \boldsymbol{\nabla}\bar{v}-g\overline{\rho^\prime w^\prime}\\+\frac{\partial}{\partial z}(\overline{\tau_x^\prime u^\prime}+\overline{\tau_y^\prime v^\prime})-\epsilon(k_e).
\end{split}
\end{equation}

\noindent where, $\rho^\prime$, $\tau^\prime_x, \tau^\prime_y$, and $p^\prime$ are the anomalies of density, wind stress (zonal, meridional), and pressure relative to the seasonal mean, and $g$ is the acceleration due to gravity. The first term of the left-hand side of the Equation \ref{eqn3} is EKE tendency, the second and third are the advection terms, and the fourth is the pressure divergence term. The first two terms on the right-hand side of the equation denote the shear production or amount of barotropic instability represented by T4 as per the LEC. This instability is associated with horizontal shear, which grows by extracting kinetic energy from the background mean flow. The expansion of these terms leads to, $-\rho_0\left[\overline{{u^\prime}^2}\frac{\partial \bar{u}}{\partial x}+\overline{u^\prime v^\prime} \left(\frac{\partial\bar{v}}{\partial x}+\frac{\partial\bar{u}}{\partial y}\right)+\overline{{v^\prime}^2} \frac{\partial \bar{v}}{\partial y}\right]$, where the contribution from $\overline{u^\prime w^\prime}\frac{\partial \overline{u}}{\partial z}$, and $\overline{v^\prime w^\prime}\frac{\partial \overline{v}}{\partial z}$ are found to be negligible. On the other hand, the third term on the right-hand side denotes the buoyancy production or baroclinic instability, which is associated with the vertical shear of the mean flow, and it converts EPE into EKE represented by the term $-g\overline{w^\prime\rho^\prime}$ (T3 as per LEC), which signifies lighter (denser) water masses are associated with upward (downward) movement. A positive value of each term implies energy conversion from the background to the perturbation eddy state. The remaining fourth and fifth terms represent the wind power and EKE dissipation rates, respectively. A flux-gradient assumption is used to define EKE dissipation rate as,

\begin{equation}\label{eqn4}
\epsilon(k_e)=\rho_0\nu\left(\frac{\partial u^\prime}{\partial z}\right)^2+\rho_0\nu\left(\frac{\partial v^\prime}{\partial z}\right)^2,
\end{equation}

\noindent where $\nu$ represents the vertical eddy viscosity, and $(\frac{\partial w^\prime}{\partial z})^2$ is negligible compared to the vertical gradients of the horizontal currents.

\subsection{Wind power}
\noindent The area-integrated wind power is given by,
\begin{equation}\label{eqn5}
WP=\int_V \frac{\partial}{\partial z}(\tau_x^\prime u^\prime+\tau_y^\prime v^\prime)dV=\int_A (\tau_x^\prime u^\prime+\tau_y^\prime v^\prime)dA.
\end{equation}

\subsection{Shear production power or barotropic energy conversion rate}
\noindent The volume integrated shear production power is given by,

\begin{equation}\label{eqn6}
SP=\int_V -\rho_0\left[{u^\prime}^2\frac{\partial \bar{u}}{\partial x}+u^\prime v^\prime \left(\frac{\partial\bar{v}}{\partial x}+\frac{\partial\bar{u}}{\partial y}\right)+{v^\prime}^2 \frac{\partial \bar{v}}{\partial y}\right] dV.
\end{equation}

\subsection{Buoyancy production power or baroclinic energy conversion rate}

\noindent The volume integrated buoyancy production power is given by,

\begin{equation}\label{eqn7}
BP=\int_V -gw^\prime\rho^\prime dV.
\end{equation}

\noindent With these definitions in hand, we now show the depth-integrated (upper 100 m) EKE in Figure S11(a) averaged over the period 20/10-21/11, along with the center of the eddy (marked by a star) as it translated westward within the chosen box ($85^\circ$E-$89^\circ$E, $14^\circ$N-$18^\circ$N). The center of the eddy has been calculated by taking the minima of the sea-level anomaly. The monthly average picture shows two anti-cyclonic eddies surrounding the eddy we are studying, and the depth-integrated EKE is high along the eddy edges. The vertical profile of the mean EKE from 84$^\circ$E-89$^\circ$E at 16.5$^\circ$N is shown in Figure S11(b). The sub-surface values of EKE are relatively high till 180 m ($\sim$75\% compared to the surface values) along the two vertical lobes of the eddy in tune with the velocity associated with the eddy in Figure \ref{f1}(g).\\ 

\noindent The spatial maps of depth-integrated values of barotropic and baroclinic conversion rates for the upper 100 m are shown in Figure \ref{f6}(a)-(h) before trapping (10/10), and during the trapping and homogenization process (specifically on 20/10, 7/11, 21/11). The barotropic conversion rates are weak but positive within the eddy, though strong negative rates are seen in Figure \ref{f6}(a), (b), which are associated with interactions across the cyclonic and anti-cyclonic eddy interface (left corner). This negative barotropic conversion rate at the eddy junction reduces in magnitude while trapping is in progress. Early on, the baroclinic conversion rate is mostly negative in and around the eddy, as shown in Figure \ref{f6}(e). But, during SSS evolution, the baroclinic conversion rates enhance along the edges of the eddy, as seen in Figure \ref{f6}(f)-(h). In terms of the LEC, these terms are depth averages of T4 (barotropic) and T3 (baroclinic), respectively. Thus, the calculations suggest that during the homogenization of SSS, both the barotropic and baroclinic instabilities are likely to have played a role in the development of the flow in and around the eddy and in the wave-like features observed in the spatial maps of the surface salinity. \\

\noindent To further understand the overall energy conversion processes, we compute the volume-integrated EKE and its tendency, area-integrated wind power (WP), volume-integrated EKE-dissipation ($\epsilon$), shear production (SP), and buoyancy production (BP) rates over the domain $85-89^\circ$E, $14-18^\circ$E and these are shown in Figure \ref{f6}(i)-(l). The volume-integrated EKE tendency (shown in blue) is negative and gradually changes its sign as we move ahead in time from 2/11 onward, which indicates that the EKE (shown in red) initially decreases within the box and then is enhanced at later stages shown in Figure \ref{f6}(i). For a short period spanning from 7/10 to 12/10, wind power was relatively high, and a modest fraction of it was dissipated (shown in red); further, with time, wind power remained low during the remaining period in consideration (Figure \ref{f6}(j)). There are instances when the dissipation slightly overshoots the wind power, and this is possibly related to other internal energy conversion pathways enhancing dissipation associated with the eddy. The volume-integrated shear production and buoyancy production are initially negative between 17/10-20/10 and changes sign (negative to positive) during the trapping and homogenization period (20/10-21/11) as shown in Figure \ref{f6}(k),(l). Thus, post-trapping, the volume-integrated T4, and T3 conversion rates are not enhanced by wind power; instead, we surmise that the entry of freshwater and eddy-mean flow interactions drives this conversion within the eddy.  \\

\noindent From the perspective of the subsurface process, three episodes are happening simultaneously: eddy-mean flow, eddy-eddy, and the eddy-freshwater interaction, which can influence shear production (T4) and buoyancy production rates (T3), simultaneously. To elucidate these processes, we show the evolution of the vertical structure of these conversion terms during the trapping and homogenization period, specifically, in the early (20/10), middle (7/11), and late (21/11) stages in Figure \ref{f7}. This also allows for understanding the evolution of the terms above and below the MLD during the homogenization process. Initially, on 20/10, the barotropic terms show a robust negative signal below the MLD down to about 80 m on the left flank of the eddy (Figure \ref{f7}(a)). This is likely due to the interaction with the neighboring anti-cyclonic eddy, as was noted in the depth-averaged view in Figure \ref{f6}(a),(b). In fact, through homogenization, most of the positive signal in T4 comes from below the MLD (Figure \ref{f7}(b)). Concerning T3, it initially shows a mixed signal about the eddy center. Going forward in time, we note a remarkable change in the baroclinic term; specifically, T3 is strongly enhanced within the mixed layer by 7/11, and as homogenization progresses, this baroclinic term within the mixed layer reduces in magnitude. Indeed, there is a clear indication that the barotropic part comes from below the MLD, where the horizontal salinity gradient is weak. At the same time, the baroclinic term is generated in the MLD (Figure \ref{f7}(e)-(f)), where the salinity stratification and strong horizontal gradients are prominent. These are correlated with the development of vertical advection of salinity ($-w\frac{\partial S}{\partial z}$) within the MLD as in Figure \ref{f7}(g)-(i). 
Thus, it appears that T4 is likely generated via eddy-eddy or eddy-mean flow interactions, while T3 is likely due to surface freshwater interaction ($T3\sim -\beta\rho_0 w^{\prime}S^{\prime}g$ in the mixed layer, where $\beta$ is the effective haline contraction coefficient of sea-water, given by $\rho^{-1}_0\frac{\partial \rho}{\partial S}$); Figure \ref{fA1}(a)-(c).  \\

\noindent In fact, this is one of the pathways observed in laboratory experiments by which ageostrophic circulation can develop in a mesoscale eddy during the adjustment of shallow, rotating freshwater vortices in a salty environment \citep{griffiths-1981,griffiths1981stability,griffiths1982laboratory}. In essence, with the help of Figure \ref{f7}, we have quantified the contribution of barotropic and baroclinic instabilities within the mixed layer during homogenization. It is to be noted in the deeper thermocline (depth $>$ 40 m) there is a development of positive buoyancy fluxes which is associated more with the temperature variations rather than salinity ($T3\sim \alpha\rho_0 w^{\prime}\theta^\prime g$ below the mixed layer, where $\alpha$ is the effective thermal expansion coefficient of sea-water, given by $-\rho^{-1}_0\frac{\partial \rho}{\partial \theta}$) as shown in Figure \ref{fA1}(d)-(f). We restrict our study within the MLD, as an instability below the mixed layer would not homogenize the freshwater. So far, we have discussed the role of instabilities but haven't considered the effect of surface buoyancy on the homogenization process. To assess this factor, we turn to the evolution of eddy available potential energy, highlighting the processes that can affect surface salinity homogenization.\\

\section{Eddy available potential energy}	
\label{epe-bud}
\noindent Eddy available potential energy (EPE) is defined by the difference of total potential energy and the minimum total potential energy resulting from any adiabatic redistribution of mass, i.e., deviation from the resting (minimum energy configuration) stable stratification state \citep{oort1989new,huang2005available,storch2012estimate}. This reads,  

\begin{equation}\label{eqn8}
P_e=-\frac{g}{2}\int_V \overline{\frac{\rho^{*\prime2}}{n_0}} dV,
\end{equation}

\noindent where $n_o$ represents the vertical gradient of the time-averaged of density ($\rho_{ref}(z)$) for the season from October to December and area-mean over a region surrounding the box $85^\circ$E-$89^\circ$E, $14^\circ$N-$18^\circ$N, respectively. The $\int_V ()dV$ in Equation \ref{eqn8} indicates the integral over the same box to a vertical depth of 100 m. $\rho^*$, $\overline{\rho^*}$ represent the deviation of $\rho$, $\overline{\rho}$ from $\rho_{ref}$, i.e.,

\begin{equation}\label{eqn9}
n_o=\frac{d\rho_{ref}}{dz},
\end{equation}

\begin{align}
&\rho^{*}=\rho-\rho_{ref}, \label{eqn10}\\
&\overline{\rho^{*}}=\overline{\rho}-\rho_{ref}, \label{eqn11}
\end{align}

\begin{equation}\label{eqn12}
\rho^{*\prime}=\rho^{*}-\overline{\rho^{*}}=\rho-\overline{\rho}=\rho^\prime.
\end{equation}

\noindent Therefore, $\rho^{*\prime}$ becomes $\rho^{\prime}$, i.e., density anomaly relative to seasonal time-mean shown in Equation \ref{eqn12}. The seasonal mean of potential temperature ($\theta$), sea-surface salinity (SSS), and potential density ($\sigma_\theta$) are shown in Figures S12(a), (b), and (c), respectively. Distinct seasonal cold water with $\theta$ (Figure S12(a)) below 29$^\circ$C, freshwater with salinity below 31 psu with $\sigma_\theta$ below 19 kg m$^{-3}$ (Figure S12(b),(c)) spans the coastal region of north-western Bay. Also, narrow river water (below 24 psu) of width $50-100$ km strip runs along the eastern coast called ``River in the Sea'' \citep{chait,shetye1996hydrography}. The vertical section of seasonal mean of $\theta$, salinity, $\sigma_\theta$, and square of Brunt-v\"{a}is\"{a}l\"{a} frequency (N$^2$) from 84$^\circ$E$-89^\circ$E at 17$^\circ$N are shown in Figure S12(d), (e), (f), (g). A temperature inversion with a very thin cold layer over the subsurface warm water (colder by 0.5$^\circ$C$-1^\circ$C) at a depth of $5-10$ m is noticeable in Figure S12(d) \citep{thad}. The salinity and $\sigma_\theta$ at the upper 15 m is of the range of $29.5-32$ psu, $\le 19.5$ kg m$^{-3}$ shown in Figure S12(e), (f). The box averaged and time-mean background $\sigma_\theta$ and its corresponding N$^2$ is shown in Figure S12(h) with $\sigma_\theta$ at the surface having the value of 19 kg m$^{-3}$. N$^2_b$ has two distinct peaks, one at $\sim$5 m, and the other subsurface peak is between $45-50$ m. These peaks are present because of near-surface freshening due to shallow river water, and the deep winter seasonal thermocline. \\

\noindent We proceed to the EPE formulation with the seasonal mean fields in place. 
Lateral gradients in potential density enhance the available potential energy of the system \citep{hughes2009available}, making it susceptible to various instabilities, thus leading to turbulence and mixing. The evolution of gradients in salinity, potential density, and EPE are shown in Figure S13. Before the freshwater entered the eddy on 3/10, the gradients in salinity and potential density, as well in the vertical section of EPE, were low (Figure S13(a), (e), (i)); as freshwater enters the eddy, these gradients build-up by lateral advection and reach their largest magnitudes around 20/10. The maximum gradients in salinity and potential density are 0.15 psu km$^{-1}$ and 0.1 kg m$^{-3}$, respectively. Comparatively, the gradient of open ocean salinity (potential density) through ship measurement in August-September of 2014 and 2015 ranges between 0.12-0.25 psu km$^{-1}$ (0.09-0.19 kg m$^{-3}$ km$^{-1}$ at 31 psu, and 29$^\circ$C) \citep{sengupta2016front}. Note that the EPE also achieves its maximum value at this time (20/10), and further, with homogenization, the EPE gradually reduces from 4/11 to 21/11. \\

\noindent The diagnostic equation of eddy available potential energy ($p_e$) is \citep{storch2012estimate},

\begin{equation}\label{eqn13}
\begin{split}
\frac{\partial p_e}{\partial    t}+\overline{\boldsymbol{u}_h}\cdot\mathbf{\nabla}_h p_e+\overline{\mathbf{u}^\prime_h\cdot \mathbf{\nabla}_hp_e}=\frac{g}{n_0}\overline{\mathbf{u}^\prime\rho^{*\prime}}\cdot\mathbf{\nabla}\overline{\rho^{*}}+g\overline{w^\prime\rho^{*\prime}}\\-g\frac{\alpha_0}{n_0}\frac{\partial}{\partial z}\overline{\rho^{*\prime}J^\prime}-g\frac{\beta_0}{n_0}\frac{\partial}{\partial z}\overline{\rho^{*\prime}G^\prime}-\epsilon(p_e).
\end{split}
\end{equation}

\noindent The first, second, and third terms on the LHS of Equation \ref{eqn13} represent the tendency and advection of EPE, respectively. The first and second terms of the RHS of Equation \ref{eqn13} represent the scalar production, i.e., conversion of MPE to EPE (T2), and the negative of buoyancy production term (-T3), respectively.  

\noindent Using Equation \ref{eqn12}, these terms reduce to,

\begin{align}
&\frac{g}{n_0}\overline{\mathbf{u}^\prime\rho^{*\prime}}\cdot\mathbf{\nabla}\overline{\rho^{*}}=\frac{g}{n_0}\overline{\mathbf{u}^\prime\rho^{\prime}}\cdot\mathbf{\nabla}\overline{\rho^{*}},\label{eqn14}\\
&g\overline{w^\prime\rho^{*\prime}}=g\overline{w^\prime\rho^{\prime}}. \label{eqn15}
\end{align}

\noindent As per the LEC \citep{lorenz1955available,boning1992eddy}, the positive value of T2 also indicates the source of baroclinic instabilities. The generation rate of EPE due to time-varying fluxes of heat and freshwater on the surface (sum of the third and fourth term on RHS of the Equation \ref{eqn13}) is denoted by $G_{p_e}$ \citep{zhan2016eddy,storch2012estimate},

\begin{equation}\label{eqn16}
\begin{split}
G_{p_e}=-g\frac{\alpha_0}{n_0}\frac{\partial}{\partial z}\overline{\rho^{*\prime}J^\prime}-g\frac{\beta_0}{n_0}\frac{\partial}{\partial z}\overline{\rho^{*\prime}G^\prime}\\
=-g\frac{\alpha_0}{n_0}\frac{\partial}{\partial z}\overline{\rho^{\prime}J^\prime}-g\frac{\beta_0}{n_0}\frac{\partial}{\partial z}\overline{\rho^{\prime}G^\prime}.\\
\end{split}
\end{equation}

\noindent where,

\begin{align}
J=\frac{Q_{net}}{\rho_0C_p}, \label{eqn17}\\
G=S_{01}(E-P), \label{eqn18}
\end{align}

\noindent and,

\begin{align}
\left(\frac{\partial \rho}{\partial \theta}\right)_{S,p}\simeq \left(\frac{\partial \bar{\rho}}{\partial \theta}\right)_{S,p} \equiv \alpha_0,\label{eqn19}\\
\left(\frac{\partial \rho}{\partial S}\right)_{\theta,p}\simeq         \left(\frac{\partial \bar{\rho}}{\partial S}\right)_{\theta,p}\equiv \beta_0. \label{eqn20}
\end{align}

\begin{equation}\label{eqn21}
Q_{net}=Q_{sw}+Q_{lw}+Q_{lat}+Q_{sen}.
\end{equation}

\noindent Here, $J, G$ represent the surface temperature and salinity fluxes and are expressed through Equations \ref{eqn17}, \ref{eqn18}. $\alpha_0$ ($=-\rho_0\alpha$) and $\beta_0$ ($=\rho_0\beta$) are values of expansion coefficients of temperature and salinity at the uppermost model layer given by Equations \ref{eqn19}, \ref{eqn20}. $Q_{net}$ and $E-P$ are the net surface heat flux and net surface freshwater flux entering the ocean. $S_{01}$ is the sea-surface salinity at 1 m, and $c_p$ is the specific heat capacity of water ($\approx$ 3940 J kg$^{-1}$ K$^{-1}$). $Q_{net}$ is the sum of incident short-wave solar radiation ($Q_{sw}$), net long-wave radiation ($Q_{lw}$), latent heat flux due to evaporation ($Q_{lat}$), and sensible heat flux due to air and water having different surface temperatures ($Q_{sen}$) in units of W m$^{-2}$ shown in Equation \ref{eqn21}.\\

\noindent For EPE analysis, as a reference, we begin with the evolution of salinity in the eddy as shown in Figure \ref{f8}(a)-(d) for early (3/10), entry of freshwater (20/10), middle (4/11), and late (16/11) stages of homogenization. Below this, we also show the evolution of volume integrated EPE to a depth of 100 m for the box 85-89$^\circ$E, 14-18$^\circ$N in Figure \ref{f8}(e), and the EPE tendency in Figure \ref{f8}(f). By comparing the movement of freshwater and EPE, we see that with the entry of freshwater, the EPE is raised and achieves the maximum value around 20/10 when the lateral gradients of salinity and potential density are maximum. 

\noindent The surface buoyancy flux ($B_0$) is expressed in units m$^2$ s$^{-3}$ reads \citep{cronin2009wind},

\begin{equation} \label{eqn22}
B_0=B_T+B_S=-g\alpha\frac{Q_{net}}{\rho c_p}+g\beta(E-P)S_0.
\end{equation}

\noindent where, $B_T$, $B_S$, are the thermal and haline components of buoyancy flux, and other variables are as defined earlier. The evolution of mean net heat fluxes ($Q_{net}$), evaporation, precipitation, evaporation minus precipitation rates ($E, P$, and $E-P$), and buoyancy fluxes ($B_T$, $B_S$, and $B_0$) at the surface of the box is shown for the period of analysis in Figure \ref{f9}(a)-(c). All of these quantities show an approximate ten-day oscillation post 20/10 (See Figure S14, i.e., the evolution of $Q_{sw}$, $Q_{lw}$, $Q_{lat}$, and $Q_{sen}$, and Figure S15 showing the H\"{o}vm\"{o}ller and latitude time diagram). During this oscillation, the minimum value of $Q_{net}$ is $\sim$ $-150$ W m$^{-2}$ suggesting that the ocean is losing heat from the surface (note, cooling makes the surface ocean less buoyant ($B_0>0$)), while the largest positive value is of 50 W m$^{-2}$ around 4/11. There were three pulses of rain of maximum $\sim$ 50 mm day$^{-1}$ over the box containing the eddy from 28/10-02/11, 8/11-12/11, and 16/11-19/11, respectively. Precipitation ($E-P>0$) makes the surface of the ocean more buoyant ($B_0<0$), and a rain rate of 5 mm day$^{-1}$ (gain in buoyancy) can be offset through loss of buoyancy by 20 W m$^{-2}$ of heat flux \citep{cronin2009wind}. Therefore, 50 mm day$^{-1}$ of rainfall more than compensates for a loss of buoyancy of 150 W m$^{-2}$ due to cooling in this period. In effect, the net vertical buoyancy at the surface shows an oscillation with a period $\sim$10 day (shown in red) with thermal ($B_T$) and haline ($B_S$) components of buoyancy in opposite phases post 23rd October in Figure \ref{f9}(c).\\

\noindent We now compute the conversion terms associated with scalar transport, negative buoyancy production, and the correlation fluxes between density anomalies due to lateral advection of freshwater and surface buoyancy. Figure \ref{f10}(a), (b) shows the volume integrated scalar production and the negative buoyancy production rates over the same box used to compute the EKE conversion terms earlier. Both the terms are high during the initial period from 3/10 to 15/10, attaining a maximum value of around 8/10. The initial increase in the EPE tendency due to scalar transport is due to the rise of wind power, which increases the near-surface eddy kinetic energy associated with advecting freshwater into the box enclosing the eddy. Further, there is a slight increase of scalar transport around 20/10, and both terms reduce in magnitude with time. In Figure \ref{f10}(c), we examine the generation rate of EPE from external sources, i.e., buoyancy; we separate the $G_{pe}$ into a sum of $HF_{pe}(=-g\frac{\alpha_0}{n_0}\frac{\partial}{\partial z}\overline{\rho^{*\prime}J^\prime})$, $FW_{pe}(=-g\frac{\beta_0}{n_0}\frac{\partial}{\partial z}\overline{\rho^{*\prime}G^\prime})$ and examine their contribution to the EPE tendency. The conversion rates ($G_{pe}$, $FW_{pe}$, $HF_{pe}$) initially were meager (till 8/10) as the density anomalies were weak; due to surface warming and lateral advection of freshwater $HF_{pe}$ increases, on the other hand, $FW_{pe}$ shows a negative sign because of $E-P$ being positive during this period, effectively we find the increase in the generation of EPE. Post 20th October, $HF_{pe}$ starts dropping and is modulated by the surface oscillation in $Q_{net}$. $FW_{pe}$ is in the opposite phase compared to $HF_{pe}$; freshwater makes the surface more buoyant, raising the EPE tendency, whereas cooling makes the surface ocean less buoyant and reduces the EPE tendency. In effect, the sum ($G_{pe}$) shows a modulation in the EPE tendency, which is given by the correlation of buoyancy flux with density anomalies and involves lateral advection of freshwater associated with surface cooling and local, regional rainfall. The resulting modulation can be seen in Figure \ref{f8}(f) post 20/10; essentially, the surface processes drive this oscillation. \\

\noindent To cross-verify the procedures, we compared the EPE tendency with the sum of scalar transport, negative buoyancy production, and the correlation fluxes of the density anomalies and surface buoyancy flux shown in Figure \ref{fA2}. It is clear that wind power enhances the EPE tendency before the freshwater gets trapped, and further, the modulation in the EPE tendency post-trapping is due to surface buoyancy processes. It is to be noted we did not calculate the mixing as we aim to understand the conversion processes during homogenization. Mixing terms and advection of EPE outside the eddy will effectively reduce the EPE tendency, which can be understood by noting the endpoint where the blue and red lines meet post-homogenization in Figure \ref{fA2}.

\section{Conclusion and Discussion}

\noindent In this work, we have presented a comprehensive analysis of trapping and homogenization of freshwater by a cyclonic mesoscale eddy on a sub-monthly timescale using an eddy-resolving ROMS simulation with a horizontal resolution of 3 km. In accordance with observations and reanalysis, upon the entry of freshwater, the simulated eddy was characterized by a shallow mixed layer (5 to 15 m), strong-salinity stratification, and inversion in temperature ($\sim$ 2$^\circ$C). The eddy was about $\sim 300$ km in diameter, with Rossby number between $0.1-0.5$, and had significant eddy kinetic energy down to about 150 m; thus, its principal structural features were also fair agreement with reanalysis data. The probability density function of surface salinity within the box of size 300 km enclosing the eddy in the early, middle, and late stages of a subseasonal period, showed an increase in the surface salinity, signifying homogenization or irreversible mixing. The trapping and subsequent homogenization of freshwater by eddies were found to occur in multiple years from 2011 to 2019, thus establishing the robustness of these processes. \\

\noindent A mixed layer salinity budget was performed, which suggested that, in addition to lateral advection, vertical advection also plays an important role during the homogenization process. As the surface salinity evolved, depth-integrated eddy kinetic energy conversion rates showed the development of both barotropic and baroclinic instabilities within the eddy. The vertical structure of these conversion rates, precisely that of the baroclinic term, had a markedly different nature above and below the mixed layer depth within the eddy. In fact, the barotropic conversion term was largest below the mixed layer and was suggestive of eddy-eddy or eddy-mean flow interactions. On the other hand, the baroclinic term became prominent during homogenization and was confined to the upper mixed layer indicating the influence of freshwater interactions. An eddy kinetic energy budget surrounding the eddy within a box was performed, and it showed that before the trapping event, the wind power was high, and a significant part of the energy was lost to dissipation. As the freshwater entered the eddy, there was a change from negative to positive volume-integrated barotropic and baroclinic conversion rates within the domain. \\

\noindent The energetics of eddy available potential energy (EPE) showed that the entrainment of freshwater in the eddy raised its EPE, and this was due to the development of strong lateral salinity and density gradients within the eddy. The EPE was lowered in time with homogenization signifying irreversible mixing. The EPE tendency also showed modulation or fluctuations associated with different time scales before and after trapping. The EPE tendency, in the beginning, was enhanced because of increased scalar transport due to an increase in wind power (for about five days), raising the near-surface EKE associated with the entry of freshwater into the box. A 10-day rainfall event happened post-trapping, which reduced incident shortwave radiation and enhanced latent heat loss, leading to further surface cooling. The scalar production term decreased, and the EPE tendency was modulated by the correlation of surface buoyancy fluxes and density anomalies associated with the lateral advection of freshwater into the eddy. Therefore, the evolution of EPE is influenced by freshwater homogenization within the eddy and surface processes in the Bay.    
As a mark of caution, it is to be noted we neither calculate the advection of EKE or EPE during their budget analysis over the box enclosing the eddy; instead, the conversion terms imply the energy transfer across different reservoirs.\\  

\noindent While the mesoscale dynamics involved in subseasonal freshwater homogenization are seen in observations \citep{paul2021eddy} and analyzed in the present numerical study, the possibility of submesoscale dynamics in winter (see Figure 9(b) of \cite{paul2021eddy} and Figure 4.23 of \cite{,paul2022stirring}) and subsequent turbulence requires more detailed, finer-scale {\it in situ} and numerical investigations within the eddy in the Bay \citep{boccaletti2007mixed,sarkar2016interplay,spiro2018submesoscale,sengupta2016front,mackinnon2016tale,ramachandran2018submesoscale,shroyer2020upper,sukhatme2020near,phillips2021progress}. In all, the pathway presented here concerning homogenization on a subseasonal scale is distinct from the seasonal changes in sea-surface salinity via vertical diffusion \citep{akhil2014modeling,wilson2016assessment} and likely plays an important role in the evolution of freshwater during the post-monsoon season along the western boundary of the BoB. \\

\makeatletter\onecolumngrid@push

\setcounter{figure}{0}
\renewcommand{\figurename}{Fig.}
\renewcommand{\thefigure}{\arabic{figure}}
	
\begin{figure*}[ht]
	\centering
	\includegraphics[width=\textwidth]{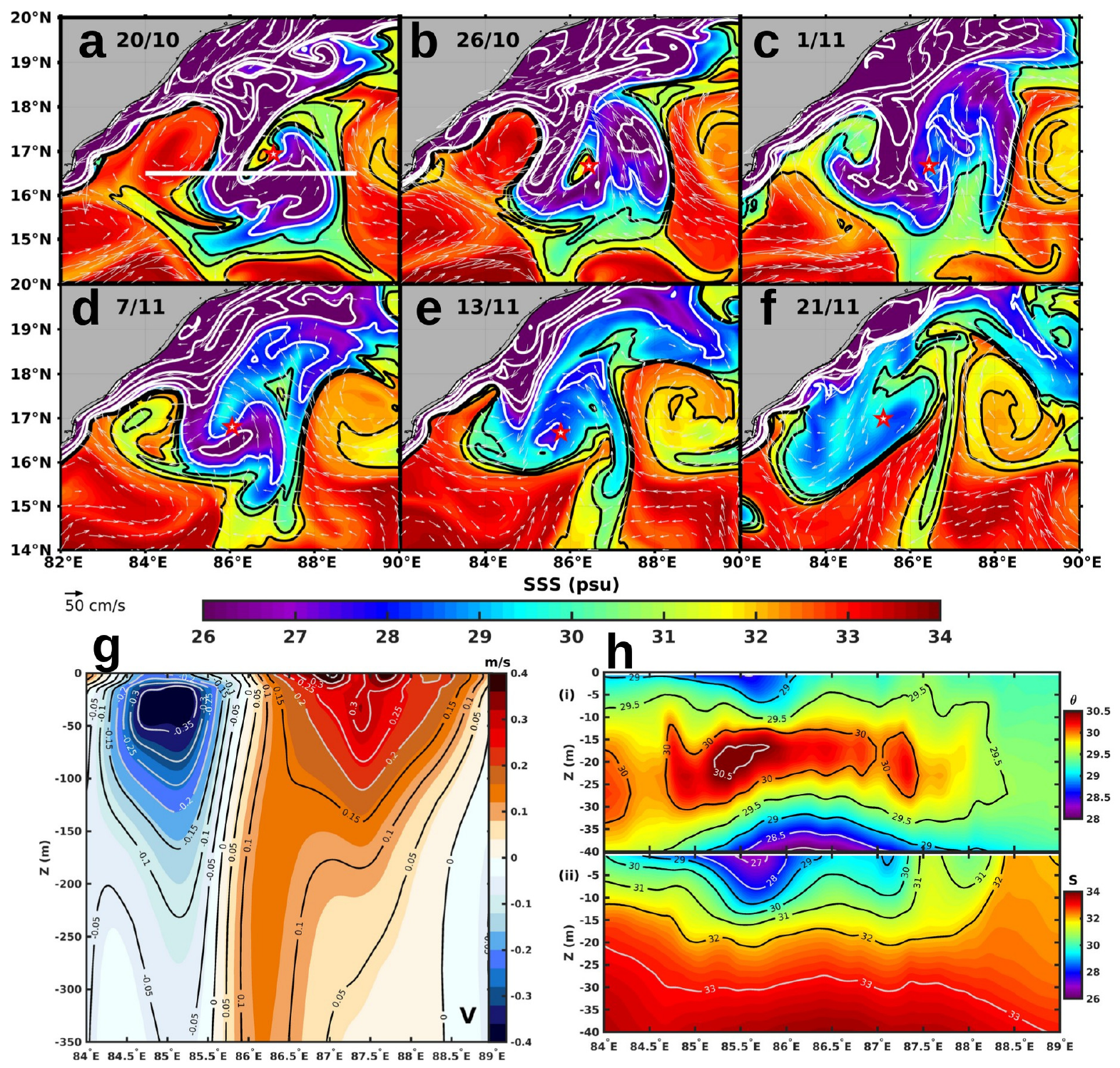}
	\caption{(a)-(f) show the evolution of sea-surface salinity with the surface current quiver (magnitude greater than 15 cm s$^{-1}$) from the model run at depths Z=-0.6 m, respectively, for the period of 20/10 to 21/11 of 2015 (dates: 20/10, 26/10, 1/11, 7/11, 13/11, and 21/11). ``Star'' marks the center of the eddy based on the minima of the sea-level anomaly. The white line in (a) marks the section of the eddy cut through 16.5$^\circ$N extending from 84$^\circ$E-89$^\circ$E. (g) Meridional velocity (m s$^{-1}$) across the cross-section of the eddy marked in panel (a) averaged over 28/10---21/11. (h) potential temperature ($^\circ$C) and salinity (psu) [(i),(ii)] averaged across the eddy section over the same period as in (g).}
	\label{f1}
\end{figure*}

\begin{figure*}[ht]
	\centerline{\includegraphics[width=\textwidth]{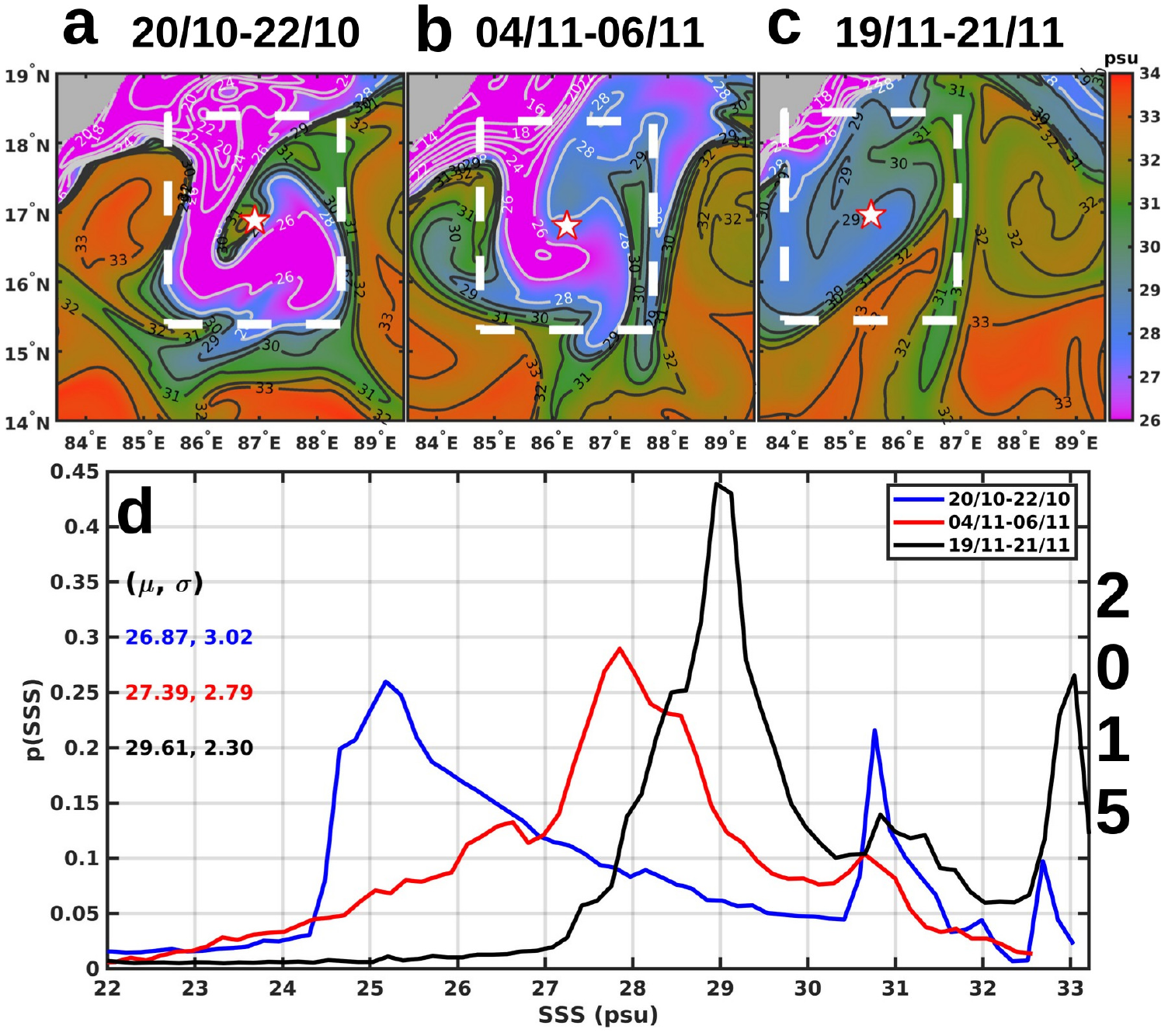}}
	\caption{(a), (b), (c) shows the average sea-surface salinity at z=-0.6 m over three-day intervals from 20/10-22/10 (early), 4/11-6/11 (middle), and 19/11-21/11 (late), of 2015 respectively, with eddy center, marked ``star'' (based on minima of sea-level anomaly). (d) The probability density function of sea-surface salinity p(SSS) over the box of size $3^\circ\times3^\circ$ about the eddy center evolving with time.}
	\label{f2}
\end{figure*}

\begin{figure*}[ht]
	\centerline{\includegraphics[width=\textwidth]{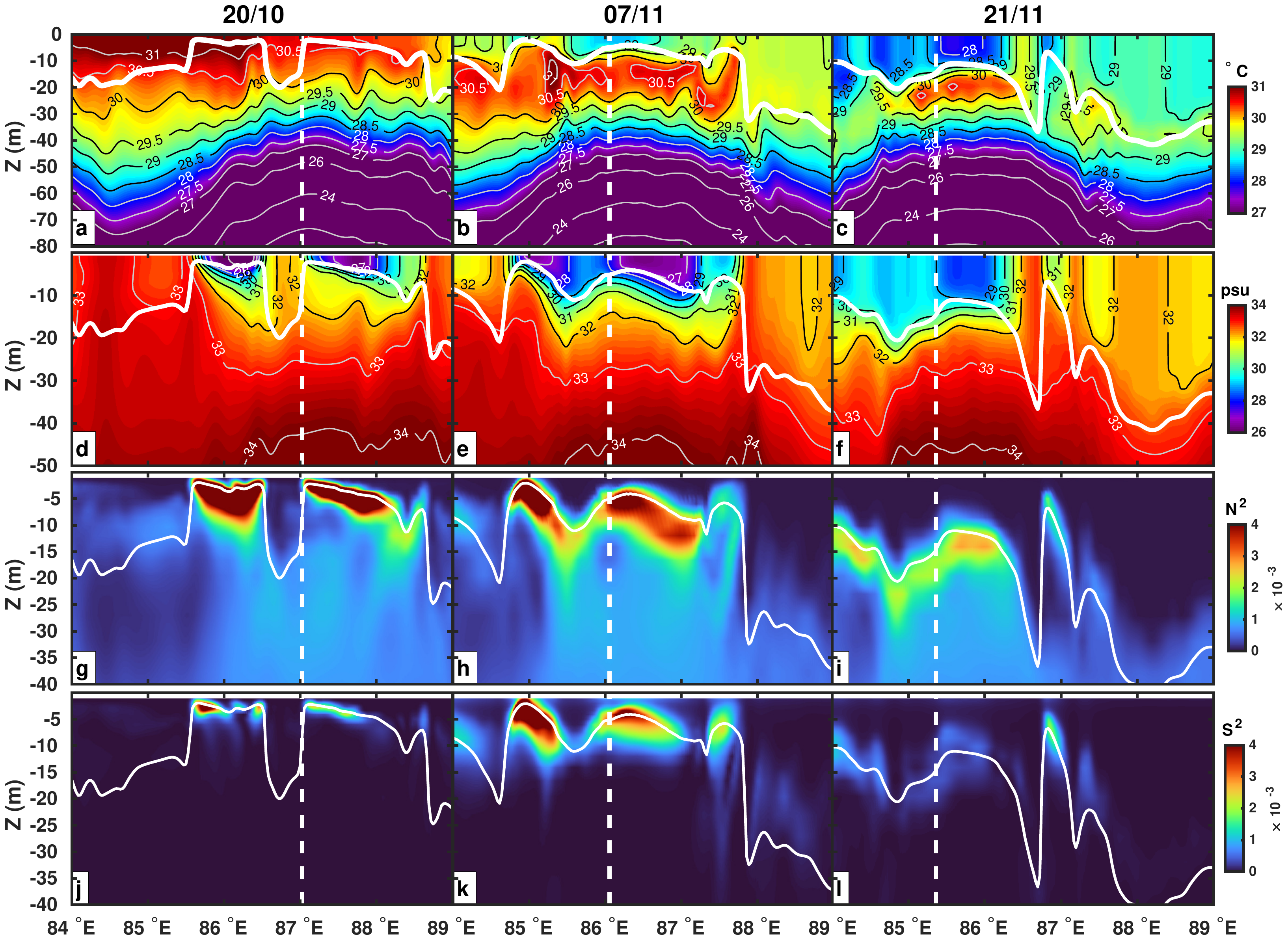}}
	\caption{(a)-(c), (d)-(f), (g)-(i), and (j)-(l) represents the potential temperature ($\theta$) [0-80 m], salinity [0-50 m], squared Brunt-v\"{a}is\"{a}l\"{a} frequency (N$^2$) [0-40 m], and square of vertical shear (S$^2$) [0-40 m] through the centre along the section of the eddy for year days 20/10, 7/11 and 21/11, respectively. The vertical dashed line in white is the reference line drawn through the center of the eddy. The white line shows the mixed layer depth through the eddy on these specific year days.}
	\label{f3}
\end{figure*}

\begin{figure*}[ht]
	\centerline{\includegraphics[width=\textwidth]{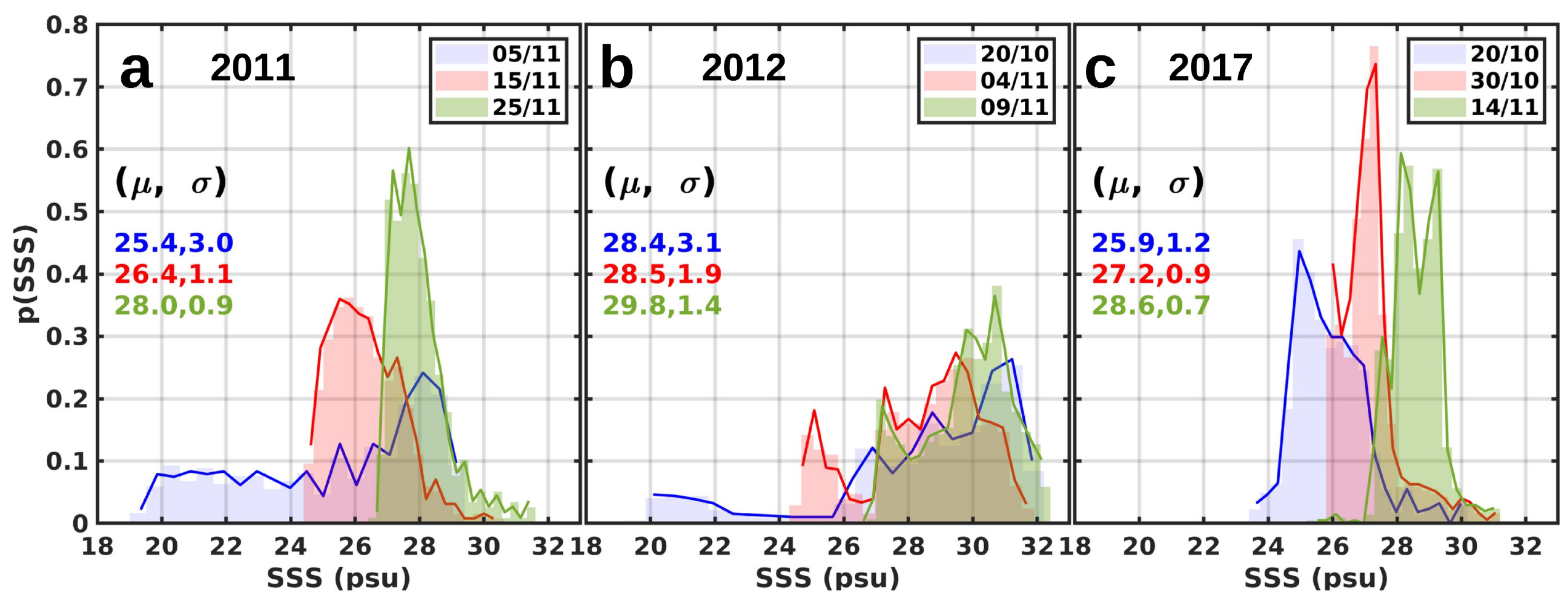}}
	\caption{The probability density function of sea-surface salinity (SSS) at Z=-0.4 m showcasing trapping and homogenization in (a) 2011 November [5/11, 15/11, 25/11], (b) 2012 October to November [20/10, 4/11, 9/11], and (c) 2017 October to November [20/10, 30/10, 14/11], within the eddy diameter of 2$^\circ$, 3$^\circ$, 2.5$^\circ$; SSS is from the Nucleus for European Modeling of the Ocean (NEMO) reanalysis. The mean ($\mu$) and standard deviation ($\sigma$) in the insets during the homogenization process.}
	\label{f4}
\end{figure*}

\begin{figure*}[ht]
	\centering
	\includegraphics[width=\textwidth]{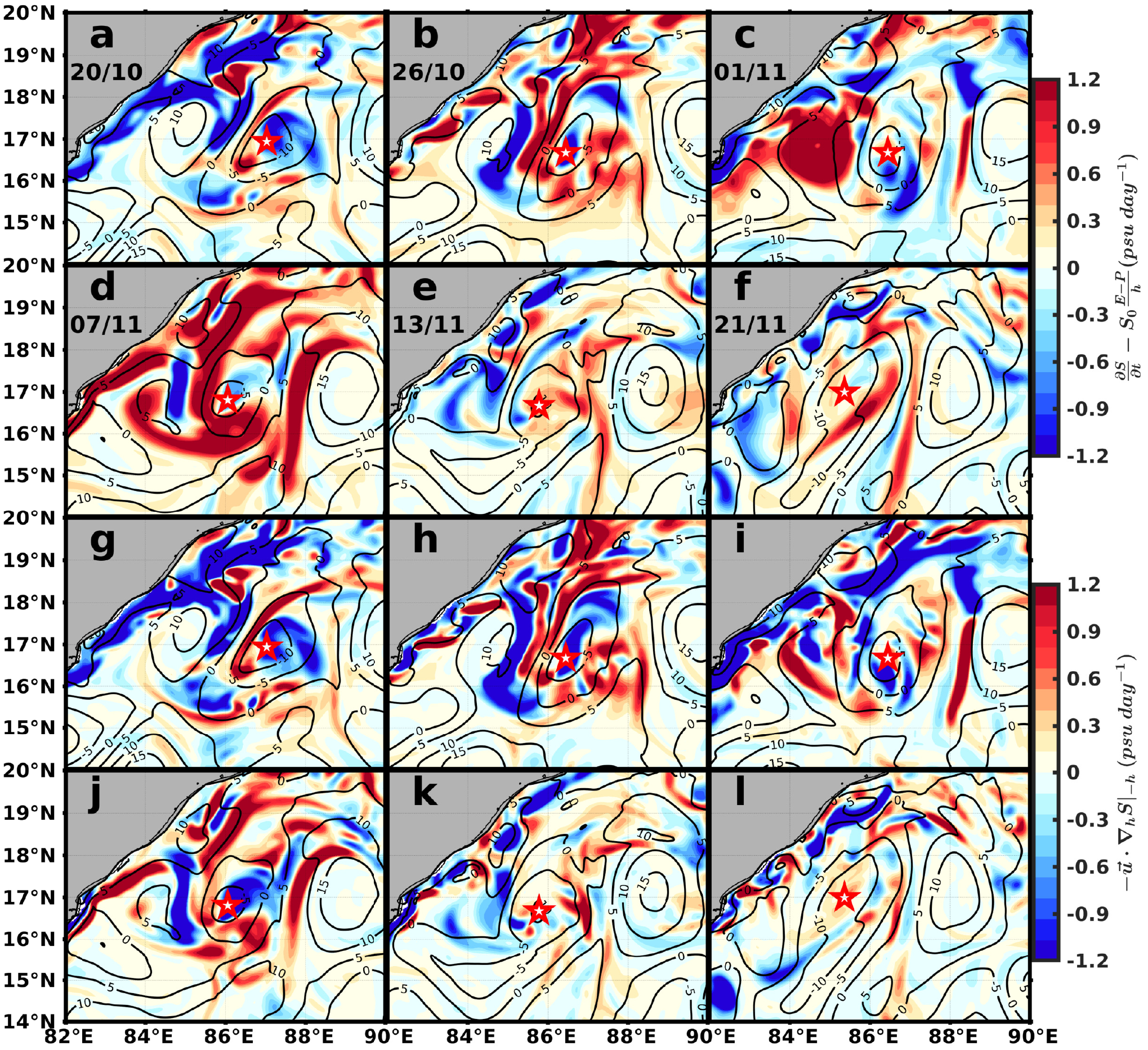}
	\caption{(a)-(f) and (g)-(l) represents the mixed layer salinity budget showing the terms $\frac{\partial S}{\partial t}-S_0\frac{E-P}{h}$ and $-\vec{u}\cdot \nabla_{h} S\vert_{-h}$ on year days 20/10, 26/10, 1/11, 7/11, 13/11 and 21/11 of 2015, respectively. ``Star'' marks the center of the eddy.}
	\label{f5}
\end{figure*}

\begin{figure*}[ht]
	\centering
	\includegraphics[height=17cm, width=16cm]{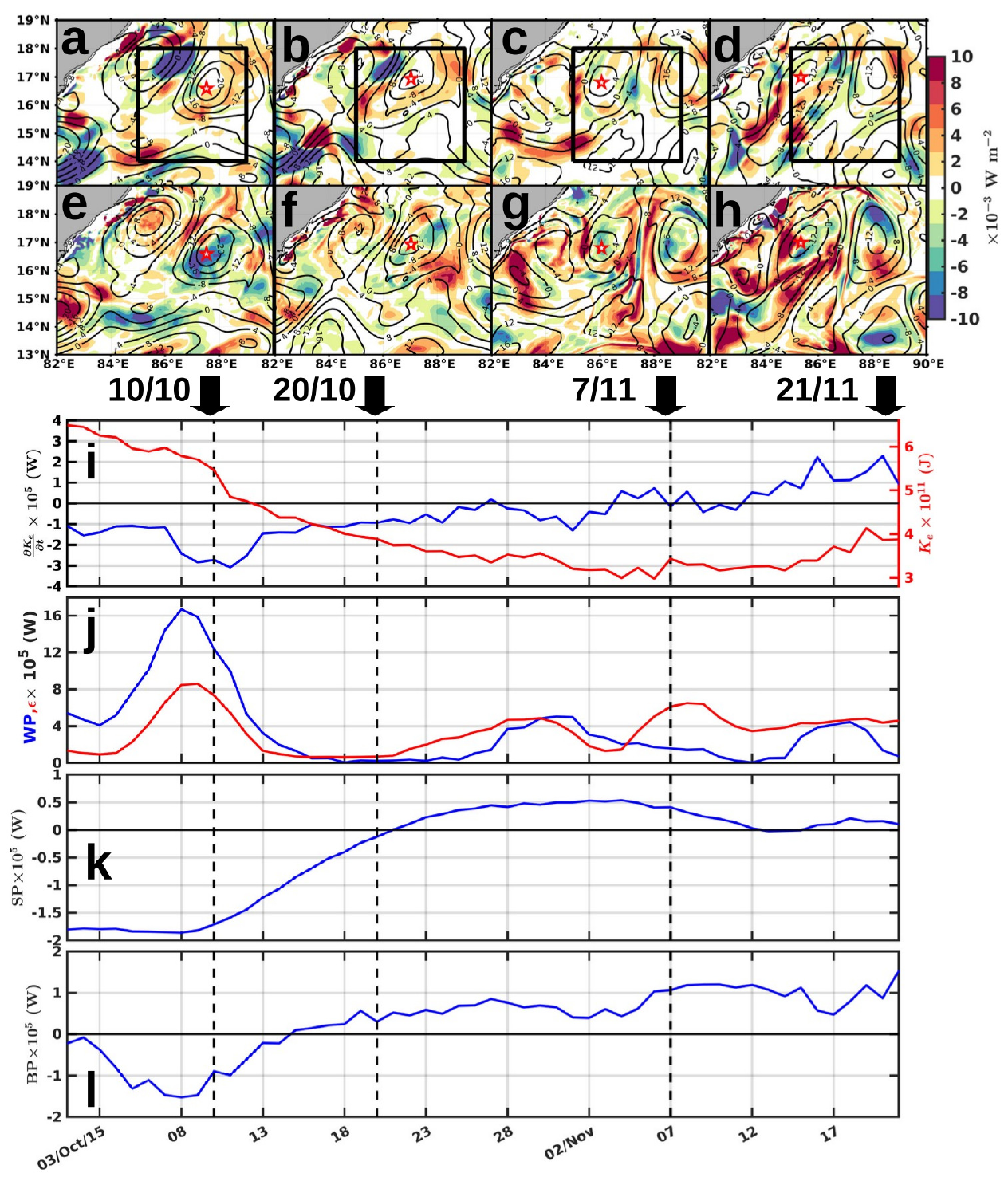}
	\caption{Depth integrated (a)-(d): shear production or barotropic energy conversion rate; (e)-(h): buoyancy production or baroclinic energy conversion rate (upper 100 m) on 10/10, 20/10, 7/11, and 21/11 of 2015, respectively. (i) Volume-integrated eddy kinetic energy (EKE) tendency, EKE in red (y-axis); (j) area-integrated wind power (WP=$\int_A (\tau_x^\prime u^\prime+\tau_y^\prime v^\prime)dA$) and volume-integrated dissipation rate [$\epsilon(k_e)=\int_V(\rho_0\nu\left(\frac{\partial u^\prime}{\partial z}\right)^2+\rho_0\nu\left(\frac{\partial v^\prime}{\partial z}\right))dV$]; (k) volume-integrated shear production (SP); (l) volume-integrated buoyancy production (BP); over 85$^\circ$E-89$^\circ$E, 14$^\circ$N-18$^\circ$N (upper 100 m). All the terms from (i) to (l) are smoothed by a 5-day moving average.}
	\label{f6}
\end{figure*}

\begin{figure*}[ht]
	\centerline{\includegraphics[width=\textwidth]{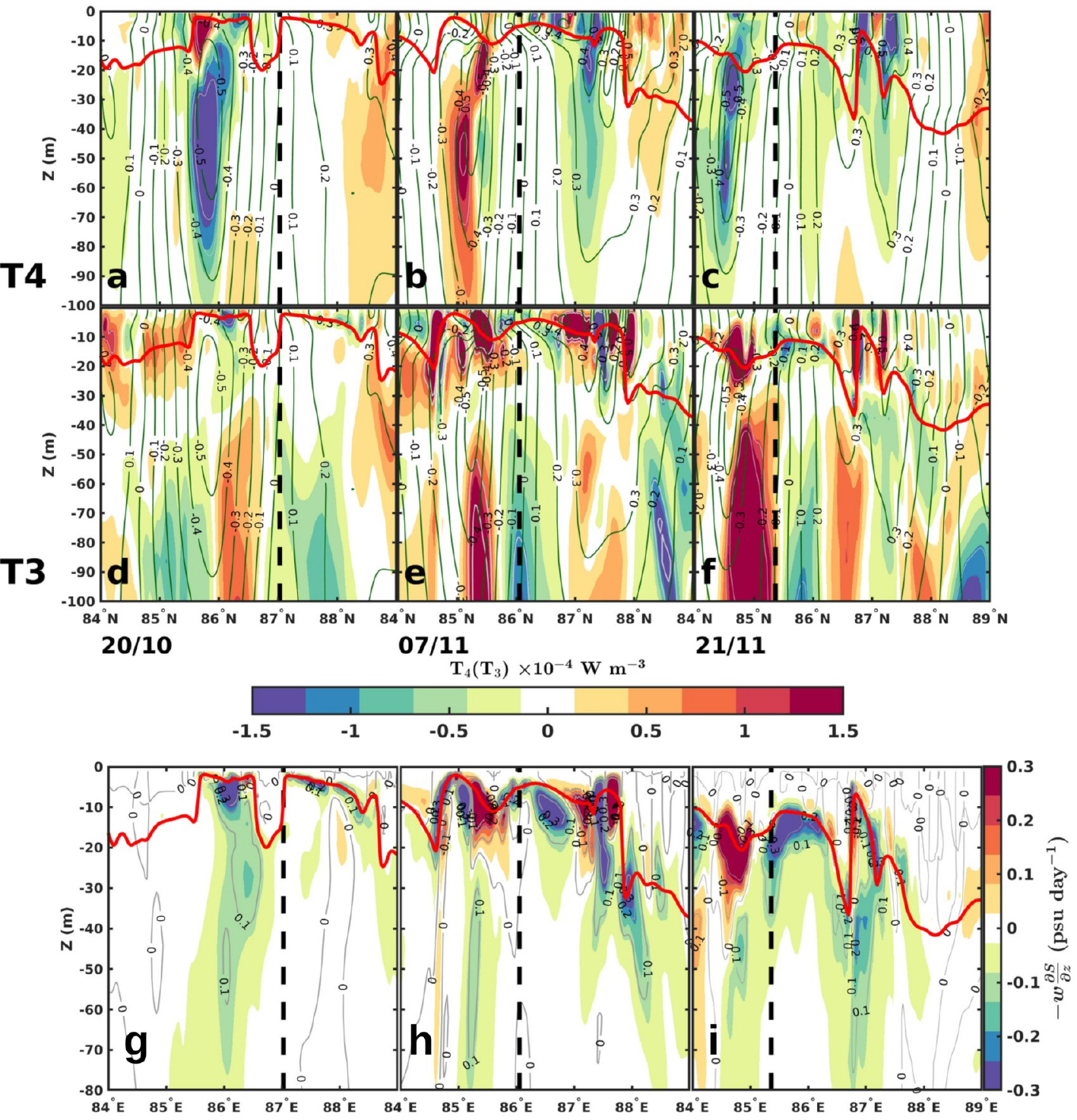}}
	\caption{(a)-(c): Vertical profile of shear production (T4), (d)-(f): buoyancy production (T3) for upper 100 m, and (g)-(i): the negative of the vertical advection of salinity ($-w\frac{\partial S}{\partial z}$) for upper 80 m with mixed-layer depth (in red) over the zonal section cut through the center of the eddy spanning from 84$^\circ$N-89$^\circ$N on 20/10, 7/11, and 21/11, respectively. The dashed line marks the center of the eddy based on the minima of the sea-level anomaly.}
	\label{f7}
\end{figure*}

\begin{figure*}[ht]
	\centering
	\includegraphics[width=\textwidth]{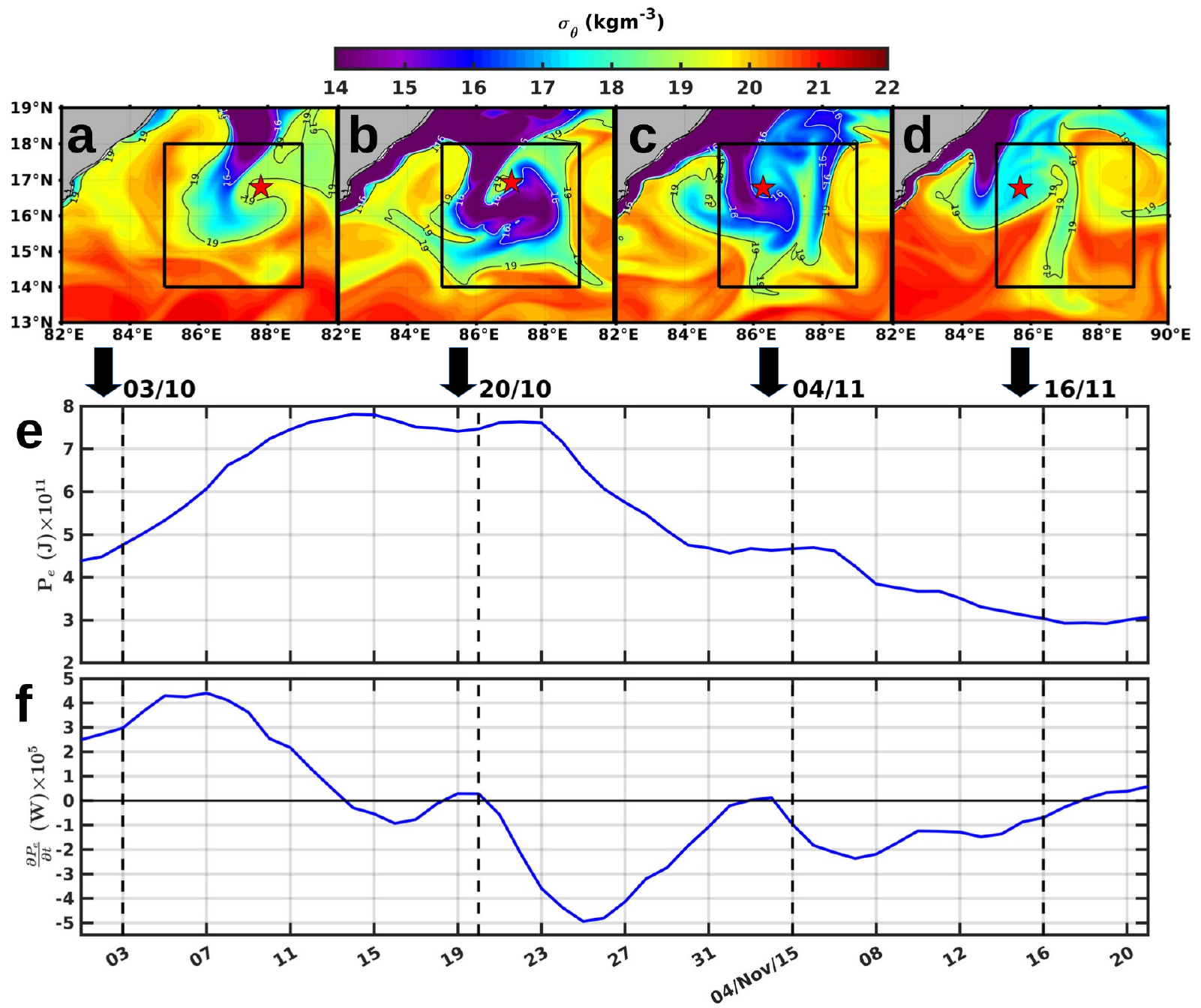}
	\caption{(a), (b), (c), (d) shows potential density ($\sigma_\theta$) during the evolution of cyclonic eddy in the box (85$^\circ$N-89$^\circ$N), 14$^\circ$N-18$^\circ$N on 3/10, 20/10, 4/11, and 16/11 of 2015, respectively. (e), (f) shows the volume-integrated available eddy potential energy $P_e$ (in Joules) and the EPE tendency $\frac{\partial P_e}{\partial t}$ (smoothed by a 5-day moving average) over the box to a depth of 100 m from 1/10 to 21/11.}
	\label{f8}
\end{figure*}

\begin{figure*}[ht]
	\centerline{\includegraphics[width=\textwidth]{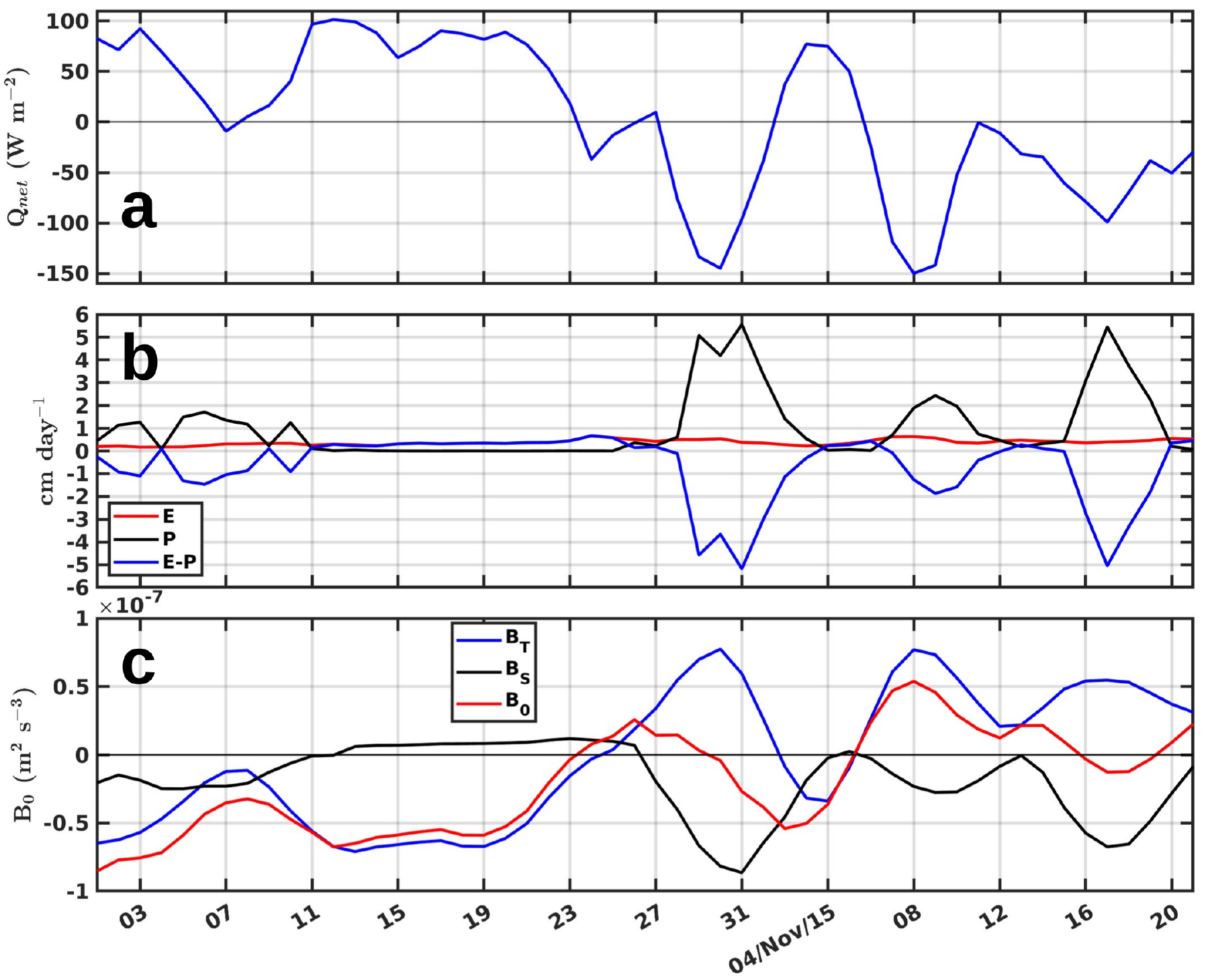}}
	\caption{(a), (b), and (c) shows the net heat radiation flux ($Q_{net}$); rate of evaporation (E), precipitation (P), and E minus P; thermal buoyancy flux ($B_T$), haline buoyancy flux ($B_S$), and buoyancy flux ($B_0$) averaged over the box: $85^\circ$E-$89^\circ$E, $14^\circ$N-$18^\circ$N shown from 1st October to 21st November of 2015. The fields $B_T$, $B_S$, and $B_0$ are smoothed by a 5-day moving average.}
	\label{f9}
\end{figure*}

\begin{figure*}[ht]
	\centerline{\includegraphics[width=\textwidth]{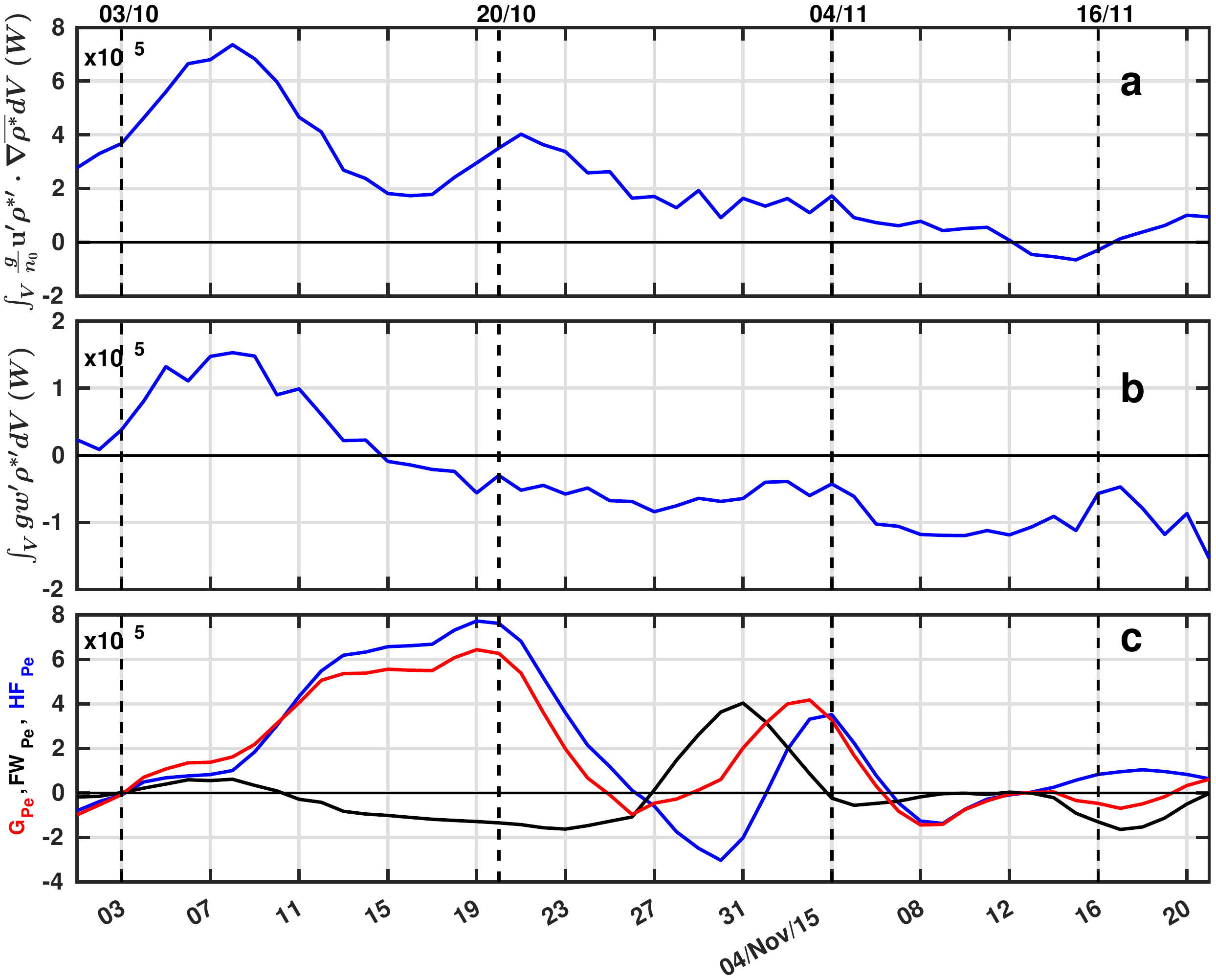}}
	\caption{(a), (b), and (c) represents the $\int_V\frac{g}          {n_0}\mathbf{u}^\prime\rho^{*\prime}\cdot\mathbf{\nabla}\overline{\rho^{*}} dV\: (W)$, $\int_Vgw^\prime\rho^\prime dV\: (W)$, and ${HF}_{Pe}=-\int_{V}g\frac{\alpha_0}{n_0}\frac{\partial}{\partial z}\rho^{*\prime}J^\prime dV\ (W)$, ${FW}_{Pe}=-\int_{V} g\frac{\beta_0}{n_0}\frac{\partial}{\partial z}\rho^{*\prime}G^\prime dV\ (W)$, $G_{Pe}={HF}_{Pe}+{FW}_{Pe}$, respectively, volume integrated over the box: $85^\circ$E-$89^\circ$E, $14^\circ$N-$18^\circ$N to a depth of 100 m shown from 1st October to 21st November of 2015. All the terms are smoothed by a 5-day moving average.}
	\label{f10}
\end{figure*}

\setcounter{figure}{0}
\renewcommand{\figurename}{Fig.}
\renewcommand{\thefigure}{A\arabic{figure}}	
 \begin{figure*}[ht]
	\centerline{\includegraphics[width=\textwidth]{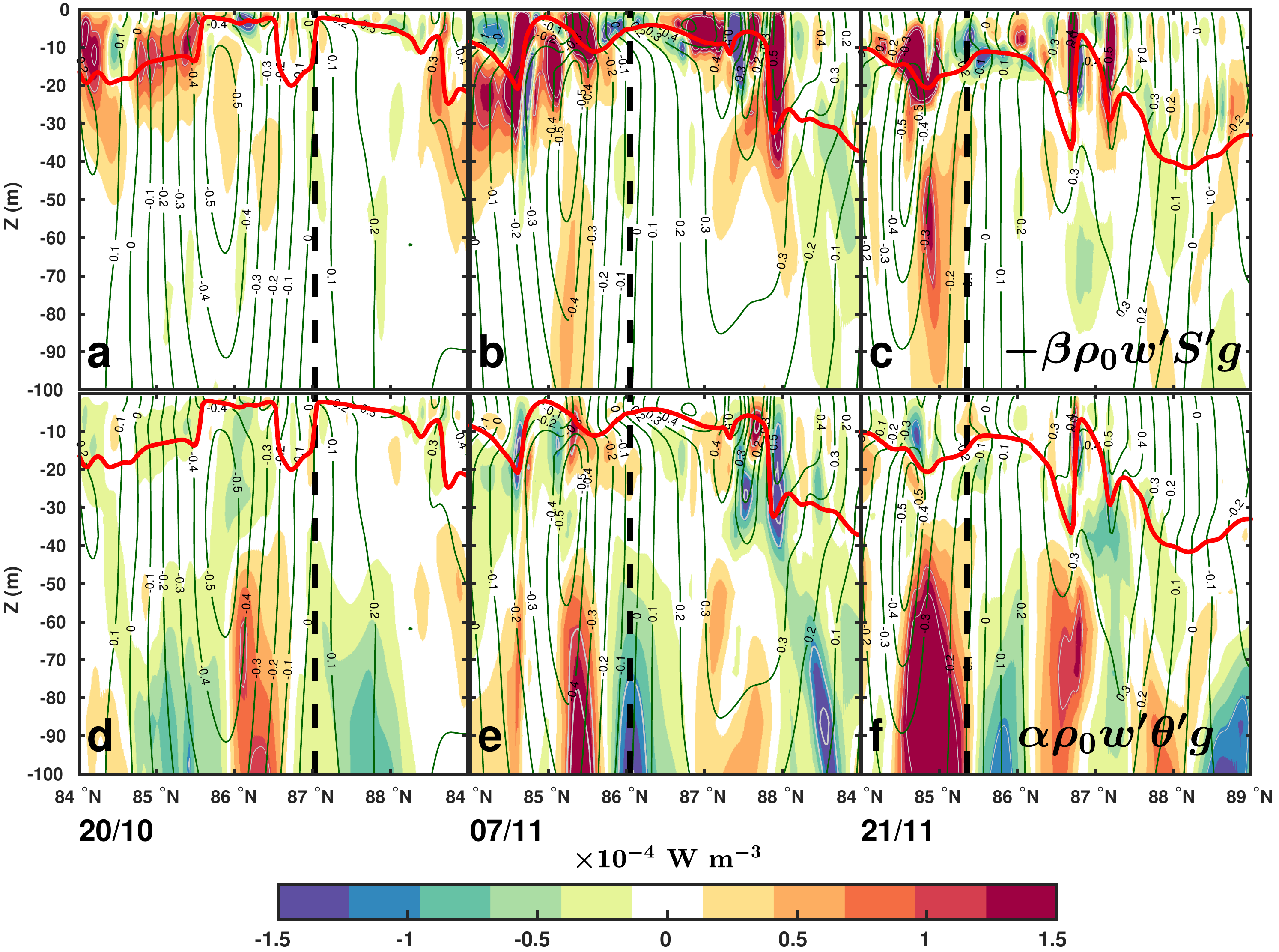}}
	\caption{(a)-(c): Vertical profile of buoyancy fluxes due to salinity ($-\beta\rho_0 S^\prime w^\prime g$), (d)-(f): buoyancy fluxes due to temperature ($\alpha\rho_0 \theta^\prime w^\prime g$) for upper 100 m with mixed-layer depth (in red) over the zonal section cut through the center of the eddy spanning from 84$^\circ$N-89$^\circ$N on 20/10, 7/11, and 21/11, respectively. The dashed line marks the center of the eddy based on the minima of the sea-level anomaly.}
	\label{fA1}
\end{figure*}

\begin{figure*}[ht]
	\centerline{\includegraphics[width=\textwidth]{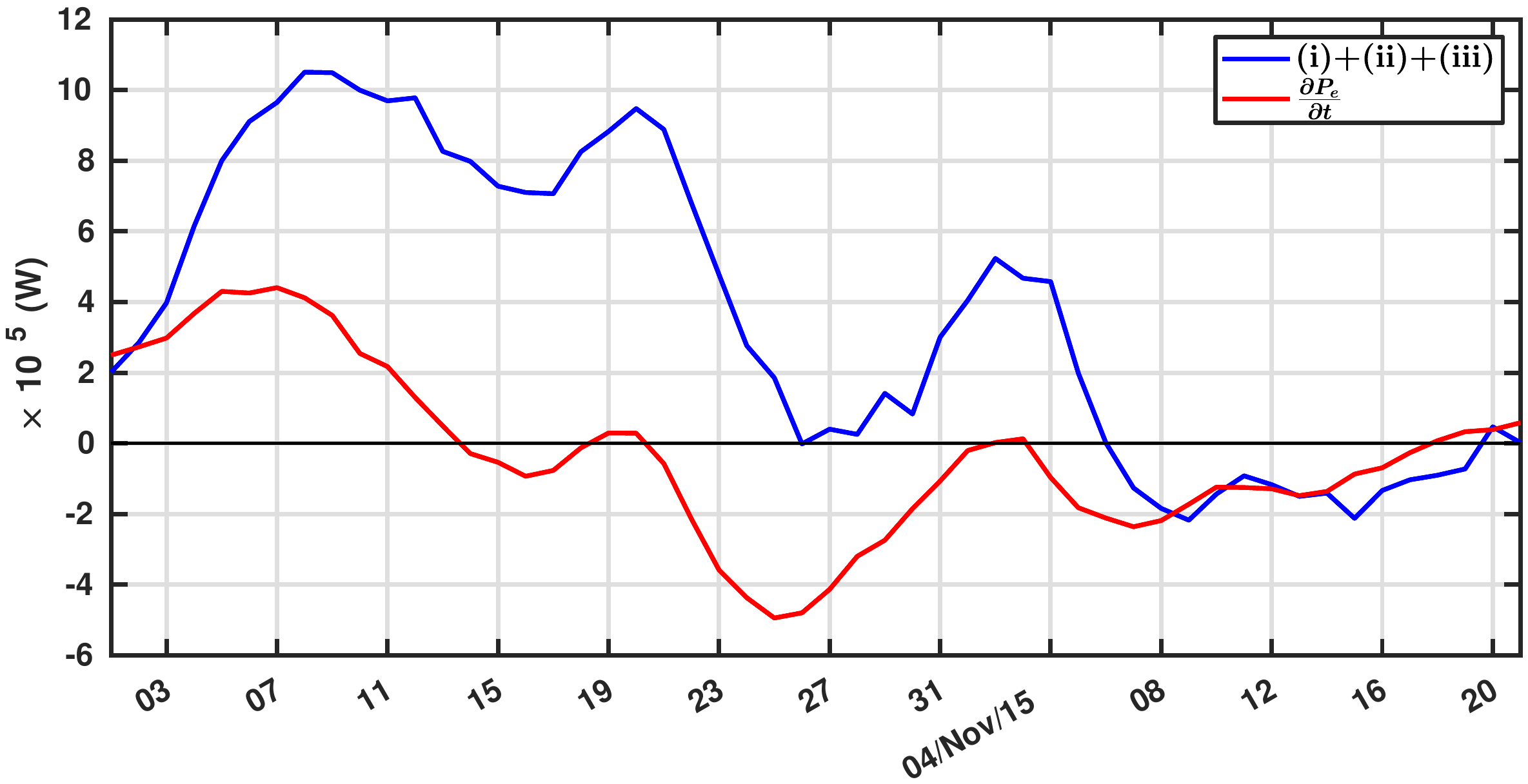}}
	\caption{The EPE tendency: $\frac{\partial P_e}{\partial t}$ (in red) vs sum of $\int_V(\underbrace{\frac{g}{n_0}\mathbf{u}^\prime\rho^{*\prime}\cdot\mathbf{\nabla}\overline{\rho^{*}}}_{(i)}+\underbrace{gw^\prime\rho^{*\prime}}_{(ii)}+\underbrace{G_{pe}}_{(iii)})dV$ (in blue) (smoothed by a 5-day moving average).}
	\label{fA2}
\end{figure*}
		
\section{Acknowledgements}
\noindent NP is grateful to Dr. Shrikant M. Pargaonkar and Dr. Subrat Kumar Mallick for discussions on Regional Ocean Modeling System. NP would also thank Dr. J. Thomas Farrar, Dr. Eric D'Asaro, and Dr. Arnold Gordon for their discussions and encouragement. The authors thank Divecha Centre for Climate Change, Indian Institute of Science, Bengaluru, India, for supporting the research work. JS acknowledges support from the University Grants Commission (UGC) under the Indo-Israel Joint Research Program (fourth cycle) for Project F 6-3/2018. DS acknowledges support from the National Monsoon Mission, Indian Institute of Tropical Meteorology (IITM), Pune, India.
	
\section{Data availability statement}	
\noindent The Nucleus for European Modeling of the Ocean (NEMO) reanalysis data (also known as GLORYS12V1 product) can be found at \url{https://resources.marine.copernicus.eu/product-detail/GLOBAL_MULTIYEAR_PHY_001_030/INFORMATION}. The surface winds at 10 m, sea-level pressure is available from \url{https://apps.ecmwf.int/datasets/data/interim-full-daily/levtype=sfc/}. The data for obtaining daily TropFlux surface fluxes and Tropical Rainfall Measuring Mission (TRMM) are as follows: \url{https://incois.gov.in/tropflux/DataHome.jsp}, and \url{https://disc.gsfc.nasa.gov/datasets/TRMM_3B42_Daily_7/summary}, respectively. The monthly climatological river discharge, World Ocean Atlas (WOA) sea-surface temperature (SST), and Soil Moisture Active Passive (SMAP) sea-surface salinity (SSS) are available on \url{https://rda.ucar.edu/datasets/ds551.0/}, \url{https://www.ncei.noaa.gov/products/world-ocean-atlas}, and \url{https://podaac.jpl.nasa.gov/SMAP}, respectively. The Regional Ocean Modeling System (ROMS) code is available at \url{https://www.myroms.org/wiki/Subversion}, and the model data used for the analysis can be found at \url{http://doi.org/10.5281/zenodo.6630083}.

\section*{Supplementary Figure}
\noindent Fig. S1 to S15

\setcounter{figure}{0}
\renewcommand{\figurename}{Fig.}
\renewcommand{\thefigure}{S\arabic{figure}}
\begin{figure}[hbt!]
	\centering
	\includegraphics[width=\textwidth]{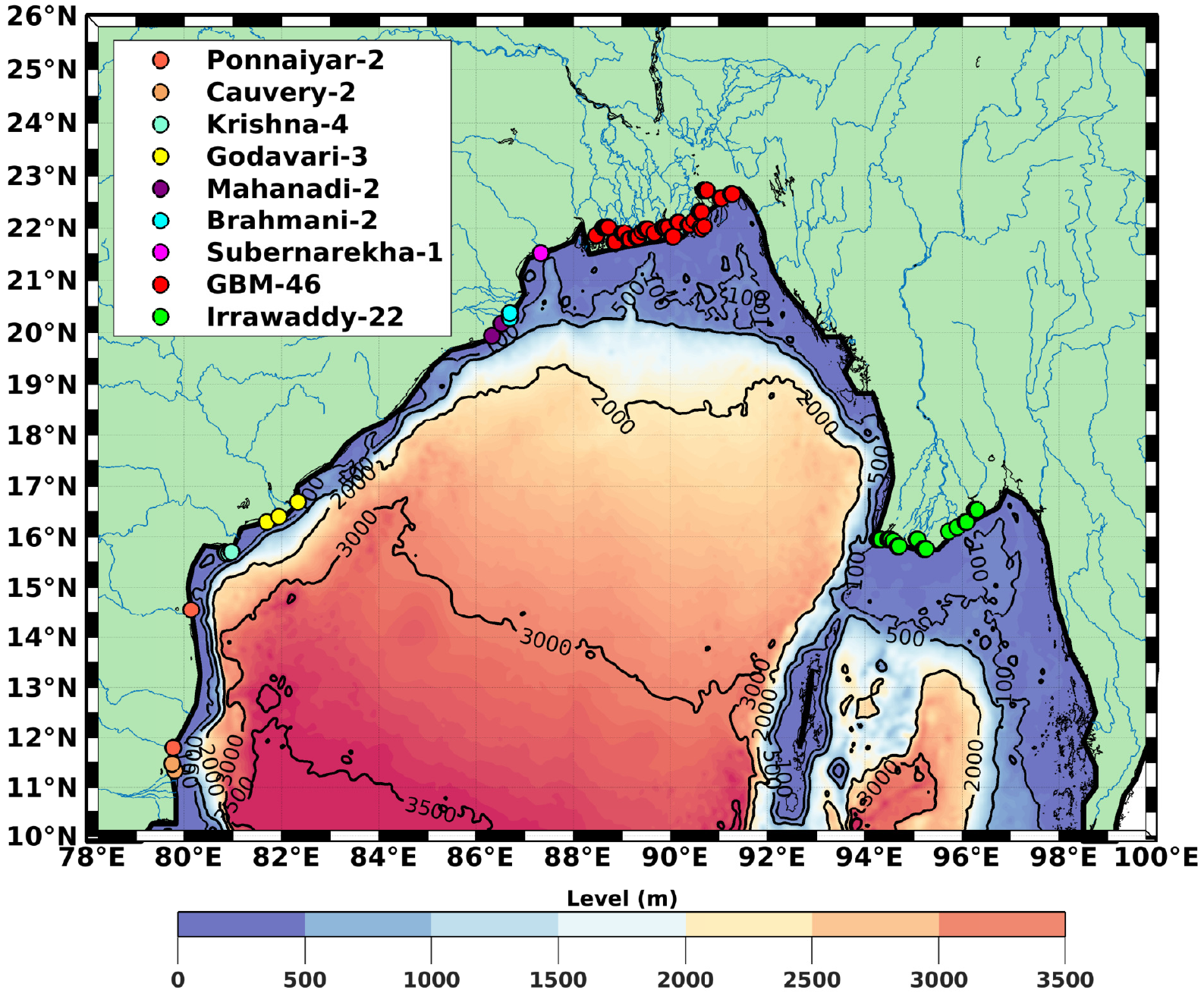}
	\caption{The bathymetry of the north Bay of Bengal (BoB) with the source points from the different tributaries in the river-BoB junction (shown in the legend) are also marked.}
	\label{s1}
\end{figure}

\begin{figure}[hbt!]
	\centerline{\includegraphics[width=\textwidth]{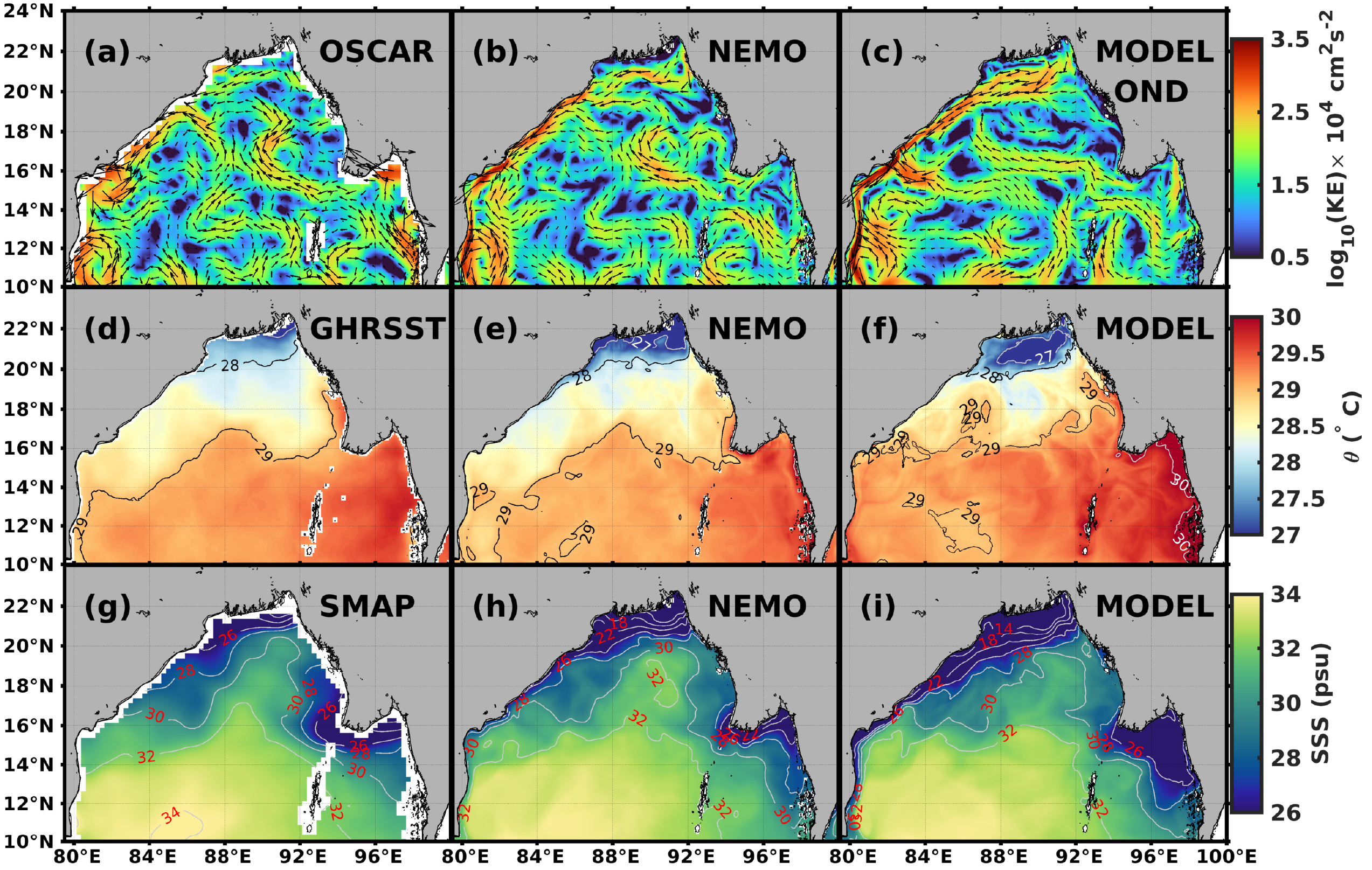}}
	\caption{October-December 2015 seasonal mean (a)-(c) kinetic energy with surface current quiver from Ocean Surface Current Analysis Real-time (OSCAR), Nucleus for European Modeling of the Ocean (NEMO) reanalysis, and model data; (d)-(f) potential temperature ($\theta$) ($^\circ$C) from Group for High-Resolution Sea Surface Temperature (GHRSST), NEMO reanalysis and model data; (g)-(i) sea-surface salinity (SSS) (psu) from Soil Moisture Active Passive (SMAP) satellite, NEMO reanalysis, and model data.}
	\label{s2}
\end{figure}

\begin{figure}[hbt!]
	\centerline{\includegraphics[width=\textwidth]{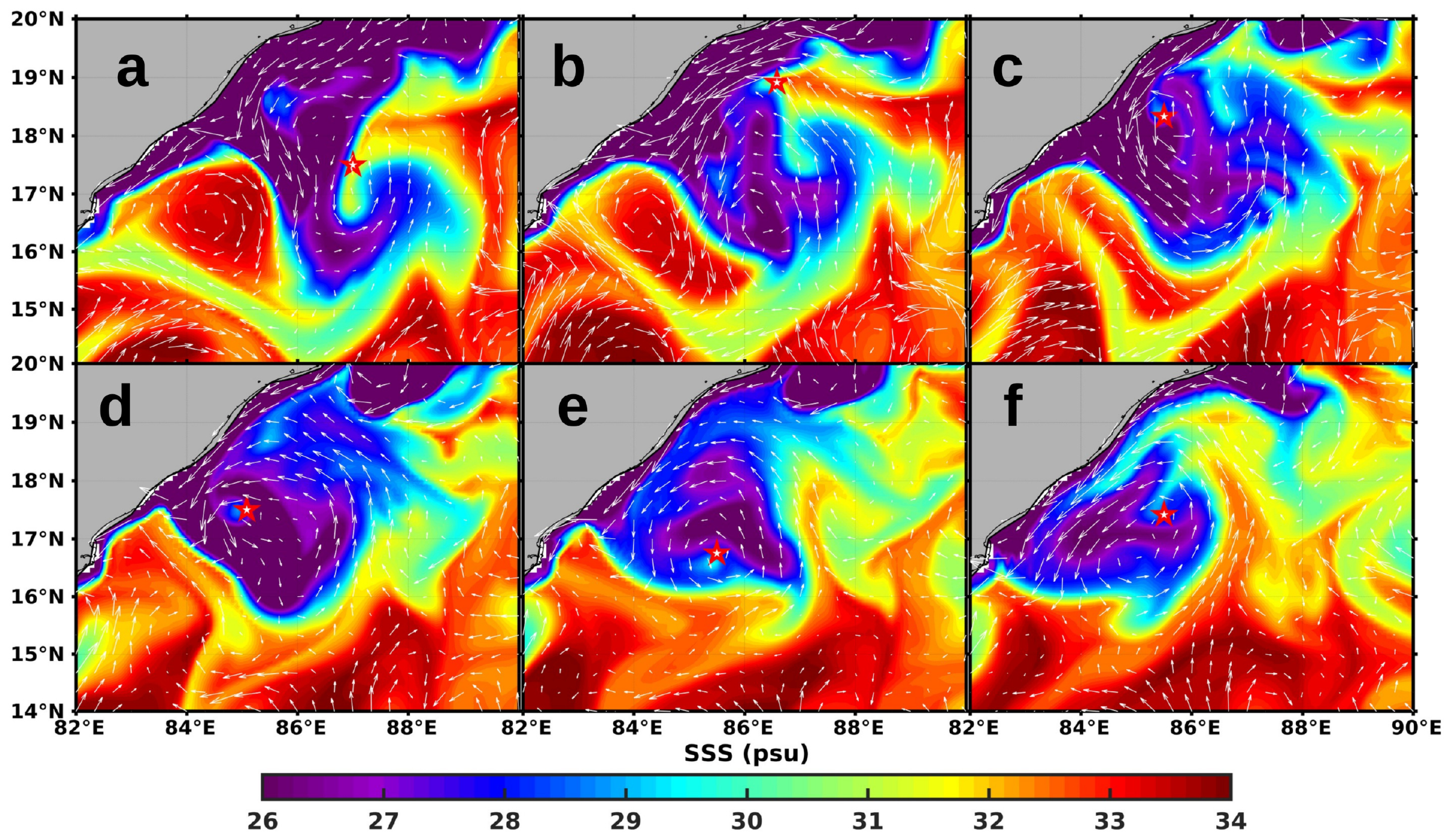}}
	\caption{(a)-(f) show the evolution of sea-surface salinity with the surface current quiver (magnitude greater than 15 cm s$^{-1}$) from the Nucleus for European Modeling of the Ocean (NEMO) reanalysis at depths Z=-0.4 m, respectively, for the period of 20/10 to 21/11 of 2015 (dates: 20/10, 26,10, 1/11, 7/11, 13/11, and 21/11). ``Star'' marks the center of the eddy based on the minima of the sea-level anomaly.}
	\label{s3}
\end{figure}

\begin{figure}[hbt!]
	\centerline{\includegraphics[width=\textwidth]{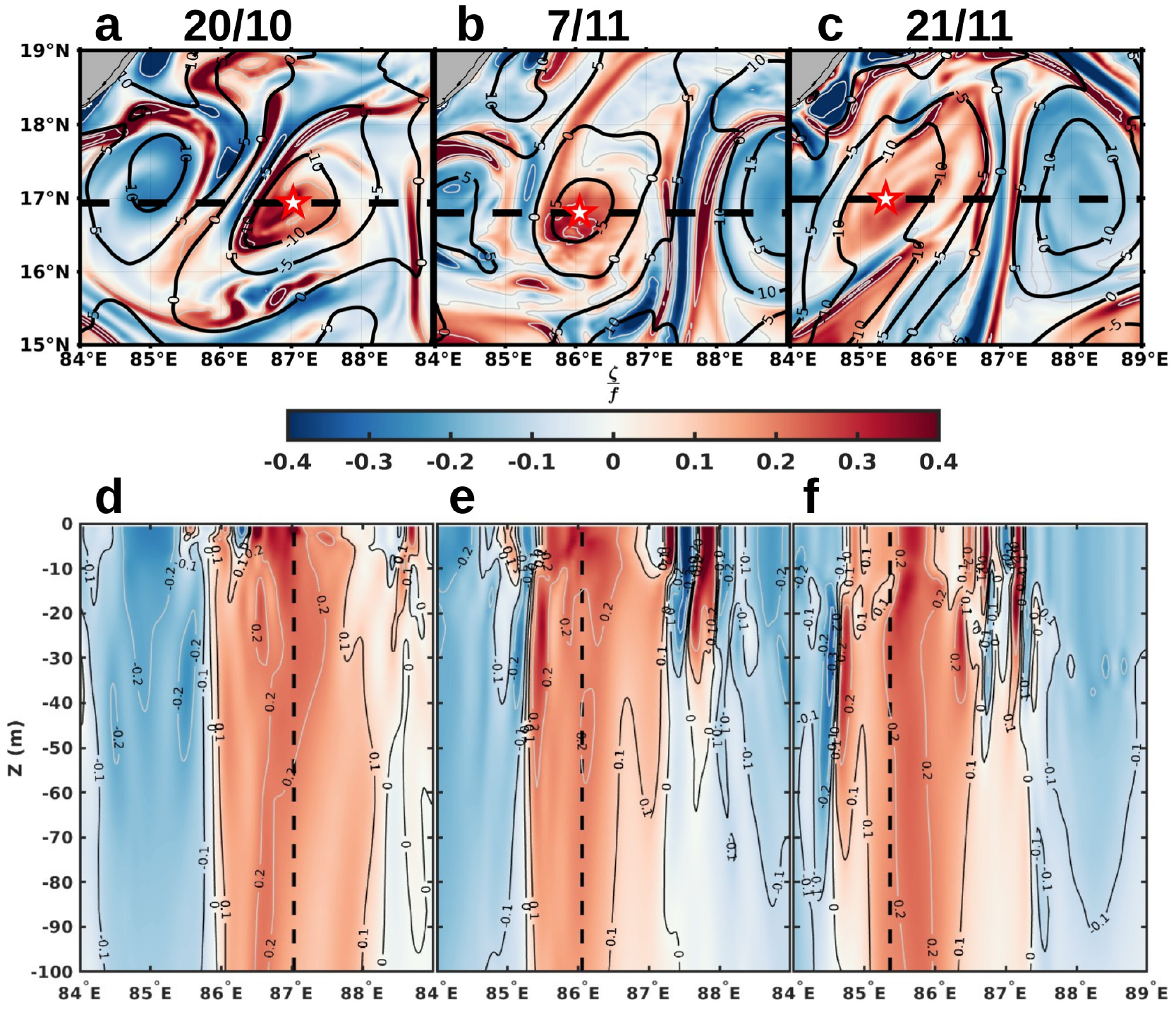}}
	\caption{(a), (b), (c) shows the Rossby number (Ro) given by $\frac{\zeta}{f}$ on 20/10, 7/11, and 21/11, respectively. (d), (e), (f) shows the vertical profile of Ro across the dashed line through the center (defined by minima of sea-level anomaly) in the eddy and is marked by ``Star''.}
	\label{s4}
\end{figure}

\begin{figure}[hbt!]
	\centerline{\includegraphics[width=25pc]{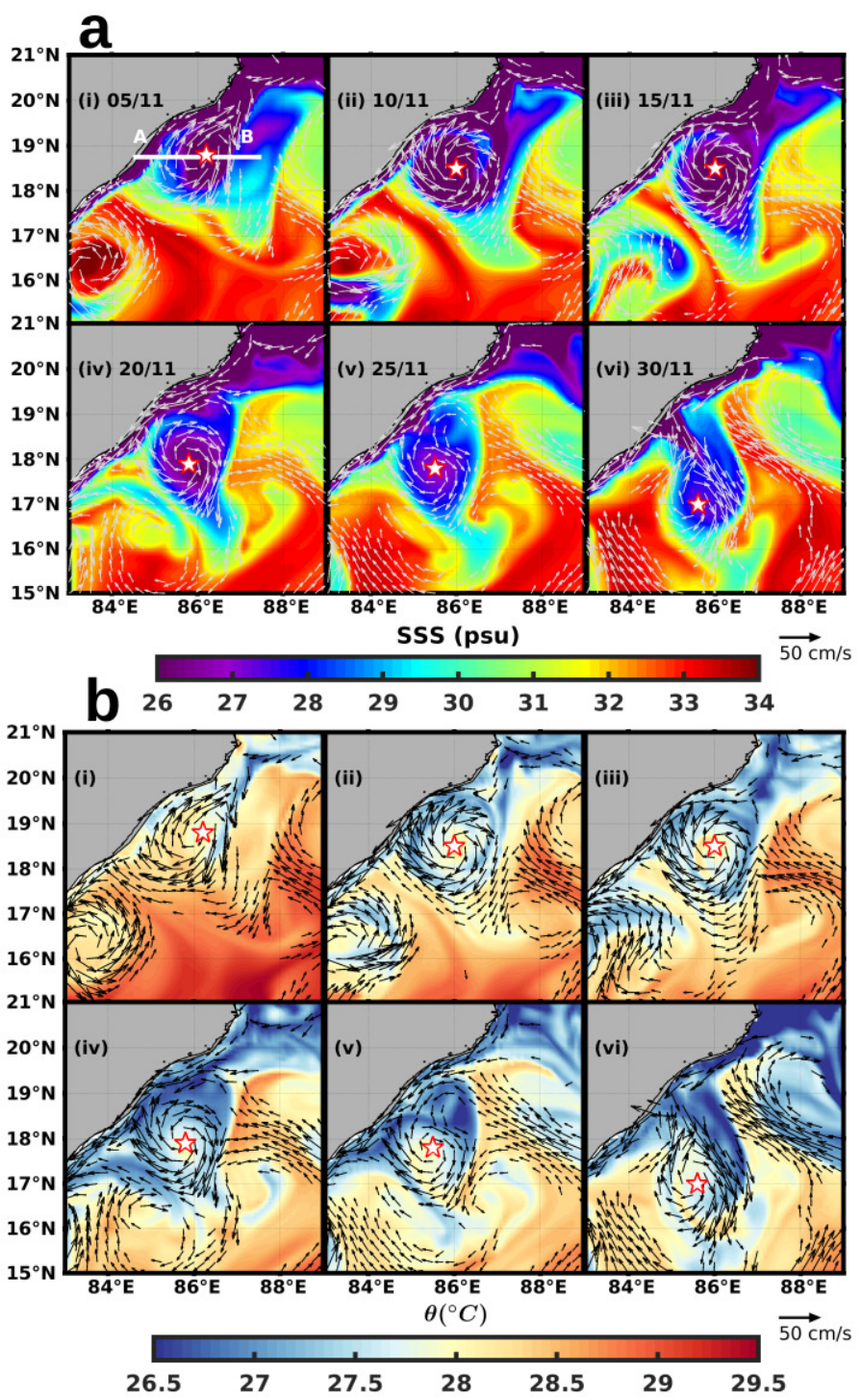}}
	\caption{(a), (b) shows the evolution of sea-surface salinity (SSS) and potential temperature ($\theta$) overlaid with a current quiver (magnitude greater than 20 cm s$^{-1}$) at z=-0.41 m from 5/11 to 30/11 of 2011 in intervals of 5 days from Nucleus for European Modeling of the Ocean (NEMO) reanalysis. Line AB denotes cross-section of the eddy in Fig. \ref{s5}(a)(i).}
	\label{s5}
\end{figure}

\begin{figure}[hbt!]
	\centerline{\includegraphics[width=\textwidth]{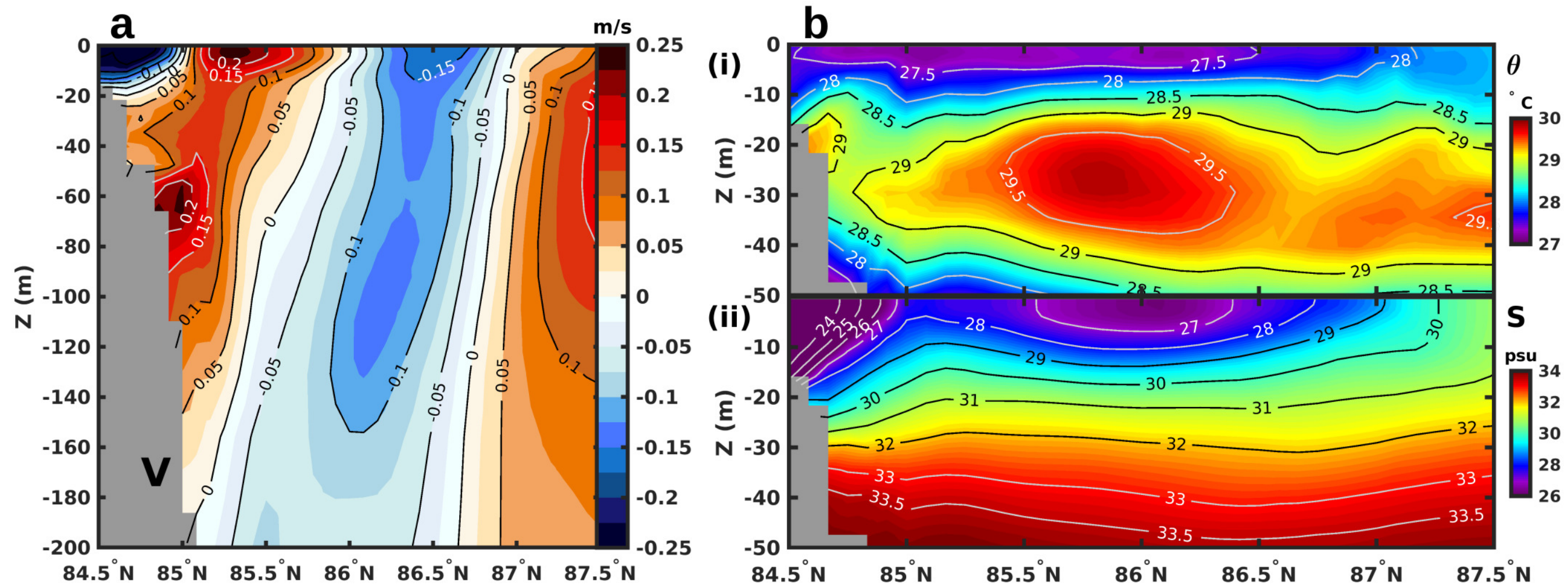}}
	\caption{(a) Meridional velocity across the longitudinal section of the eddy marked in Fig. \ref{s5}(a)(i) averaged from 5/11/2011---30/11/2011. (b) potential temperature ($^\circ$C) and salinity (psu) [(i), (ii)] averaged across the eddy section over the same period as in panel (a).}
	\label{s6}
\end{figure}

\begin{figure}[hbt!]
	\centerline{\includegraphics[width=25pc]{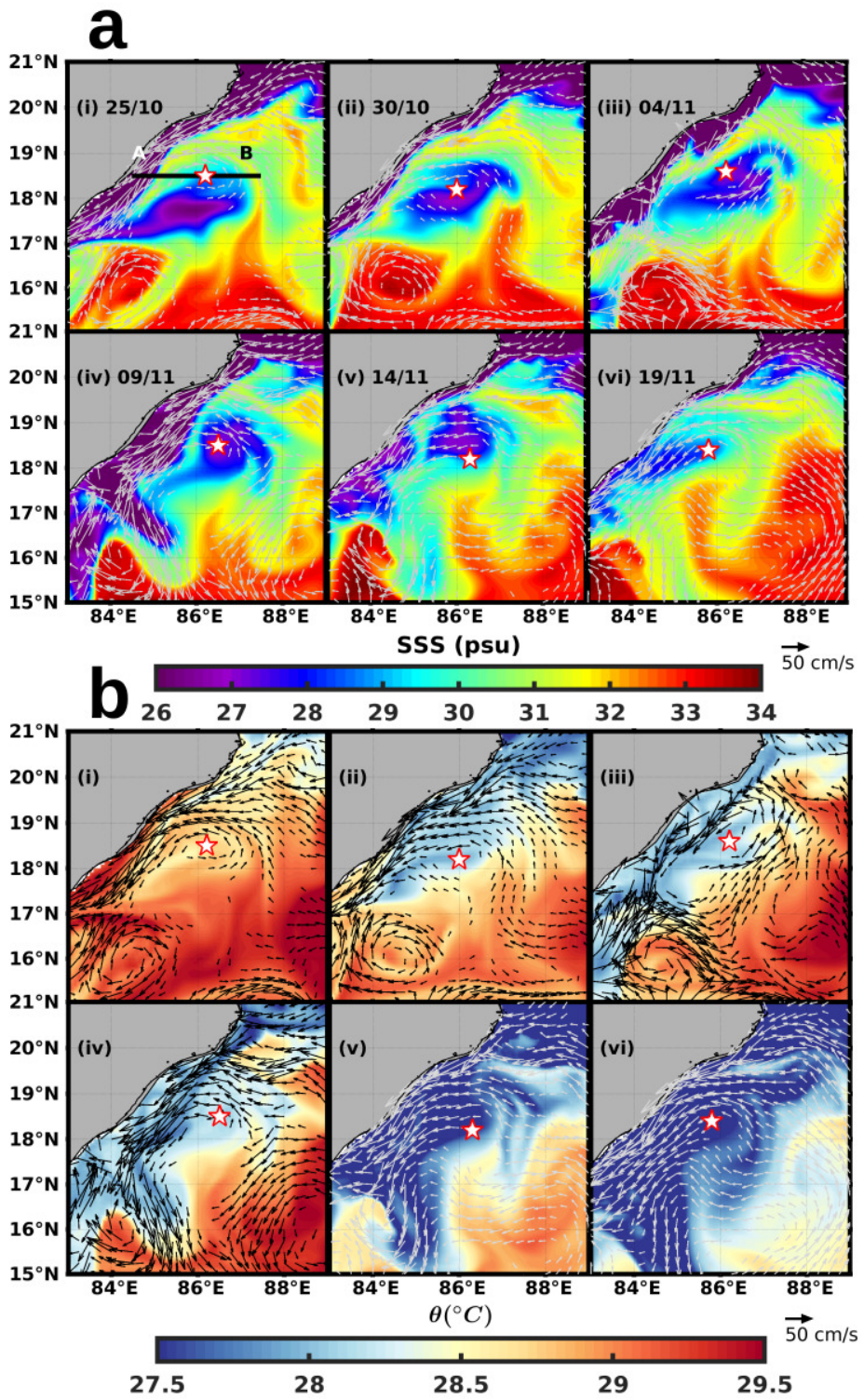}}
	\caption{(a), (b) shows the evolution of sea-surface salinity (SSS) and potential temperature ($\theta$) overlaid with the current quiver (magnitude greater than 10 cm s$^{-1}$) at z=-0.41 m from 25/10 to 19/11 of 2012 in intervals of 5 days from Nucleus for European Modeling of the Ocean (NEMO) reanalysis. Line AB denotes cross-section of the eddy in Figure \ref{s7}(a)(i).}
	\label{s7}
\end{figure}

\begin{figure}[hbt!]
	\centerline{\includegraphics[width=\textwidth]{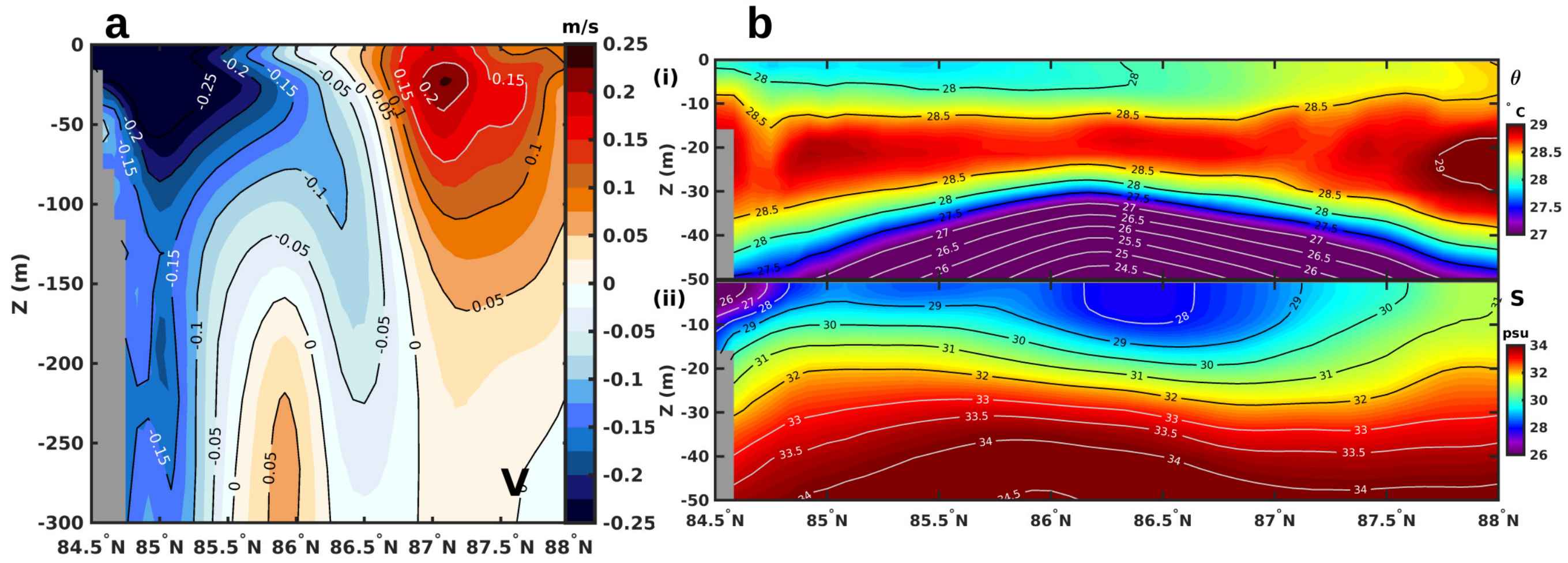}}
	\caption{(a) Meridional velocity (V) along the cross-section of the eddy marked in Figure \ref{s7}(a)(i) averaged over 25/10---19/11 of 2012. (b) potential temperature ($^\circ$C) and salinity (psu) [(i), (ii)] averaged across the eddy section over the same period as in panel (a).}
	\label{s8}
\end{figure}

\begin{figure}[hbt!]
	\centerline{\includegraphics[width=25pc]{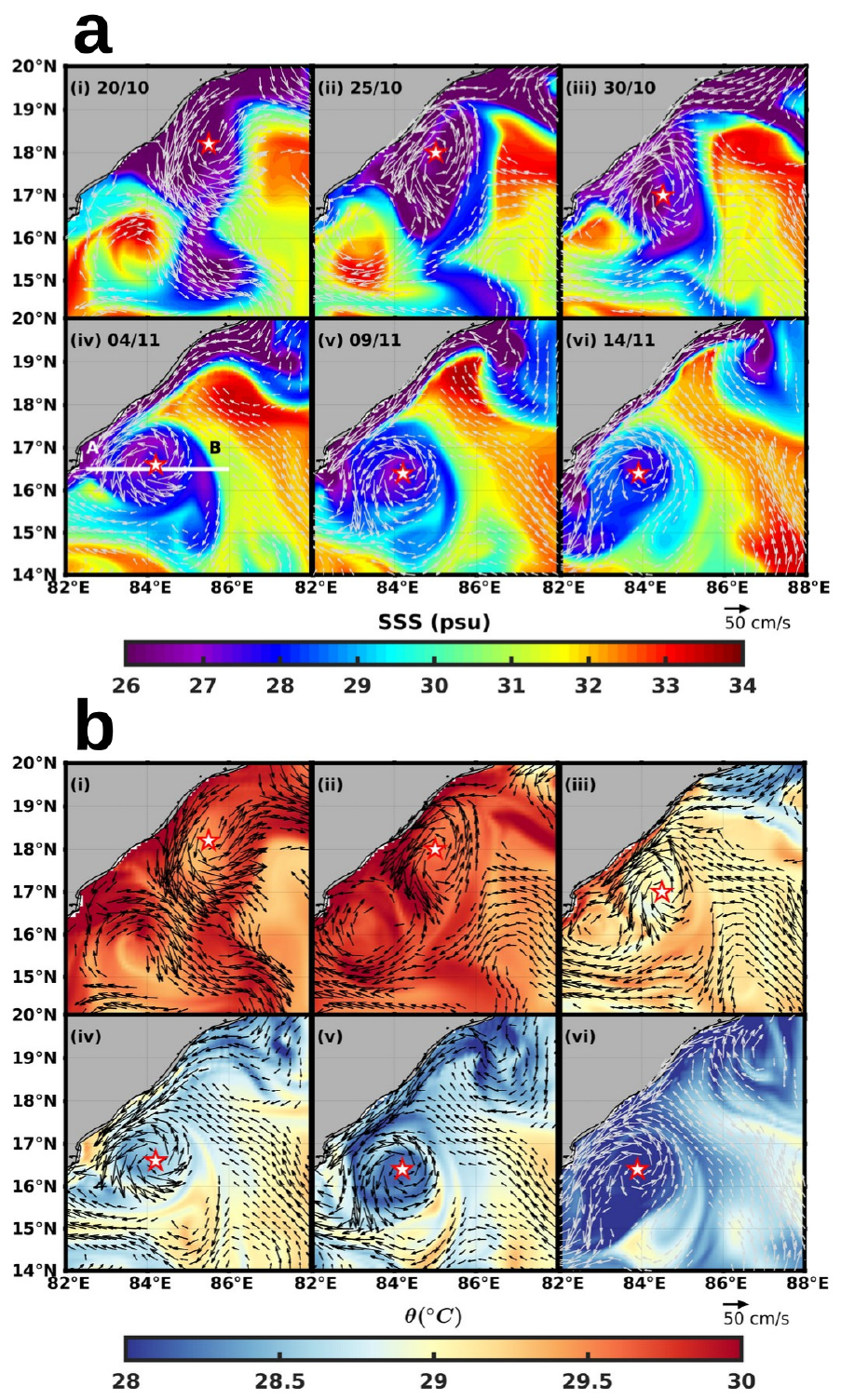}}
	\caption{(a), (b) shows the evolution of sea-surface salinity (SSS) and potential temperature ($\theta$) overlaid with the current quiver (magnitude greater than 20 cm s$^{-1}$) at z=-0.41 m from 20/10 to 14/11 of 2017 in intervals of 5 days from Nucleus for European Modeling of the Ocean (NEMO) reanalysis. Line AB denotes cross-section of the eddy in Figure \ref{s9}(a)(iv).}
	\label{s9}
\end{figure}

\begin{figure}[hbt!]
	\centerline{\includegraphics[width=\textwidth]{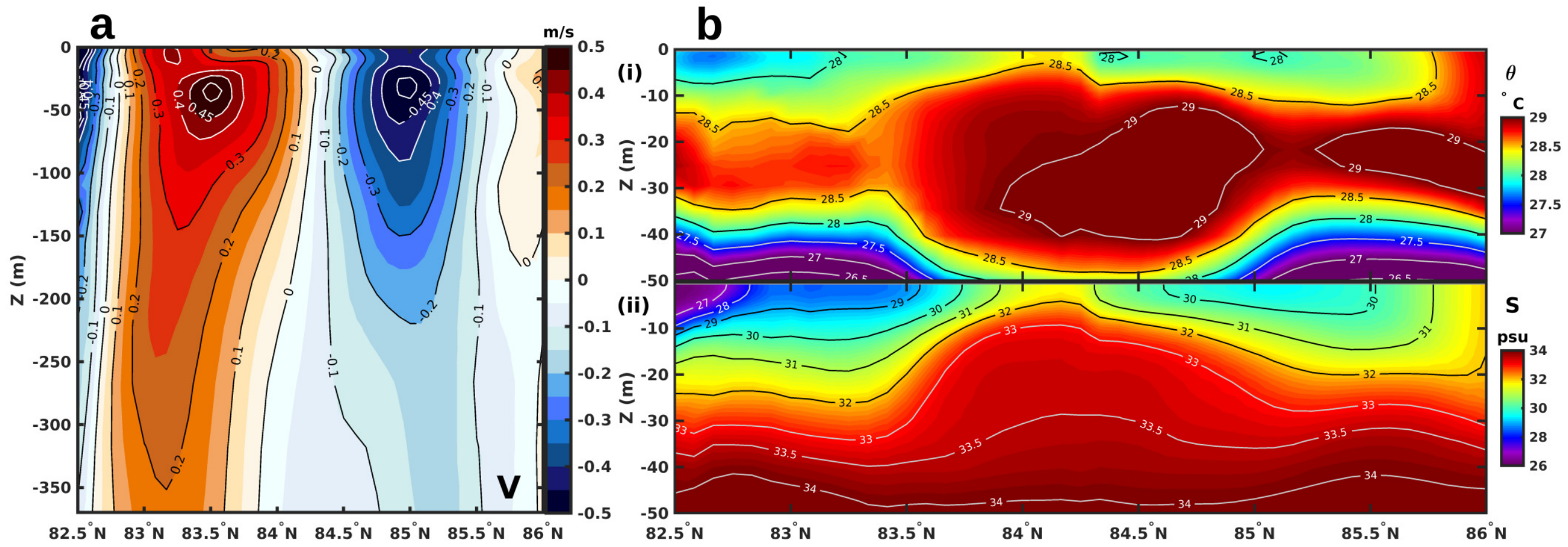}}
	\caption{(a) Meridional velocity (V) across the cross-section of the eddy marked in Fig. \ref{s9}(a)(iv) averaged over 20/10---14/11 of 2017. (b) potential temperature ($^\circ$C) and salinity (psu) [(i),(ii)] averaged across the eddy section over the same period as in panel (a).}
	\label{s10}
\end{figure}

\begin{figure}[hbt!]
	\centerline{\includegraphics[width=\textwidth]{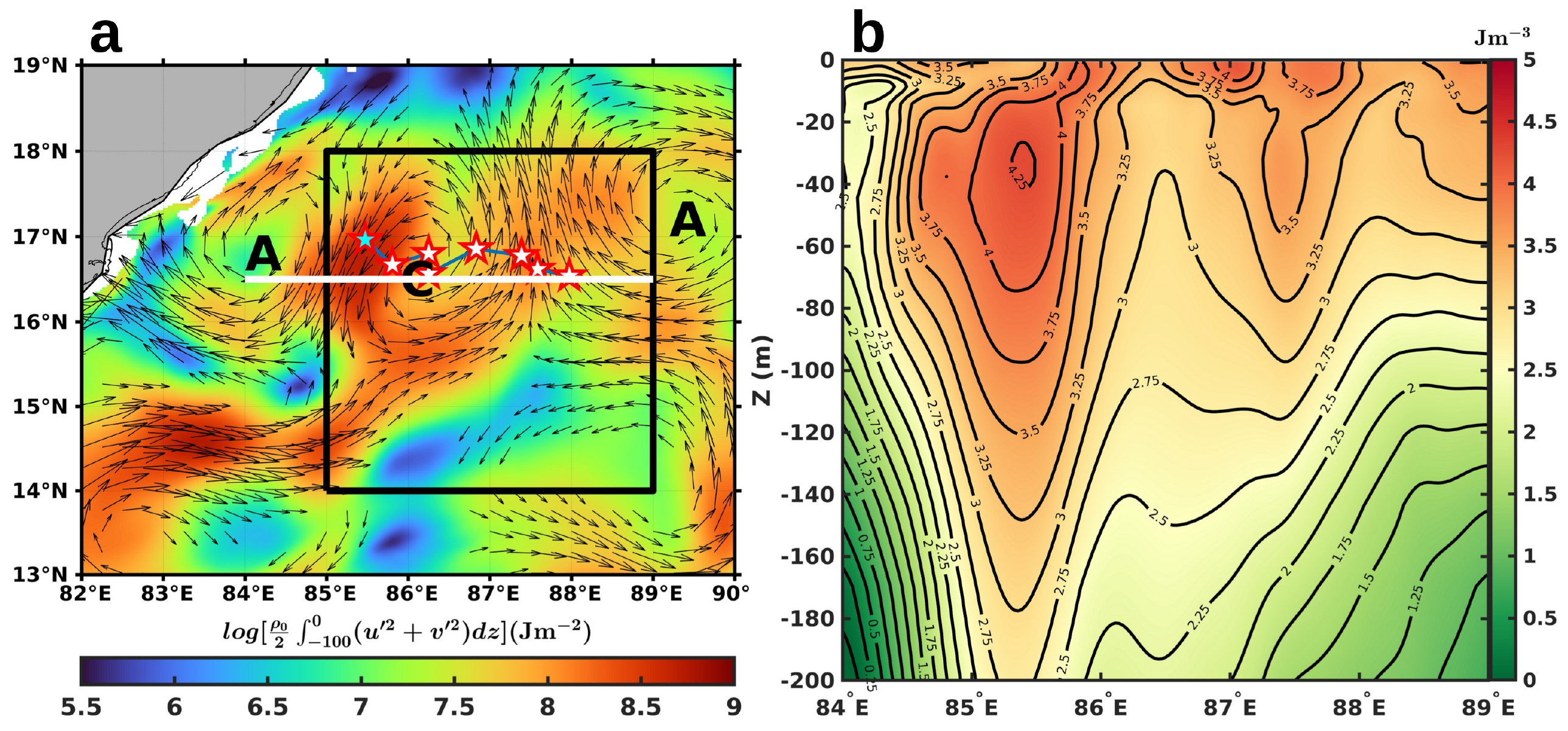}}
	\caption{(a) Depth-integrated EKE (upper 100 m); (b) EKE along 16.5$^\circ$N along the line shown in panel (a). A---Anti-cyclonic eddy, C---Cyclonic eddy. ``Star'' marks the eddy center from 1/10 to 21/11 (7-day interval).}
	\label{s11}
\end{figure}

\begin{figure}[hbt!]
	\centerline{\includegraphics[width=\textwidth]{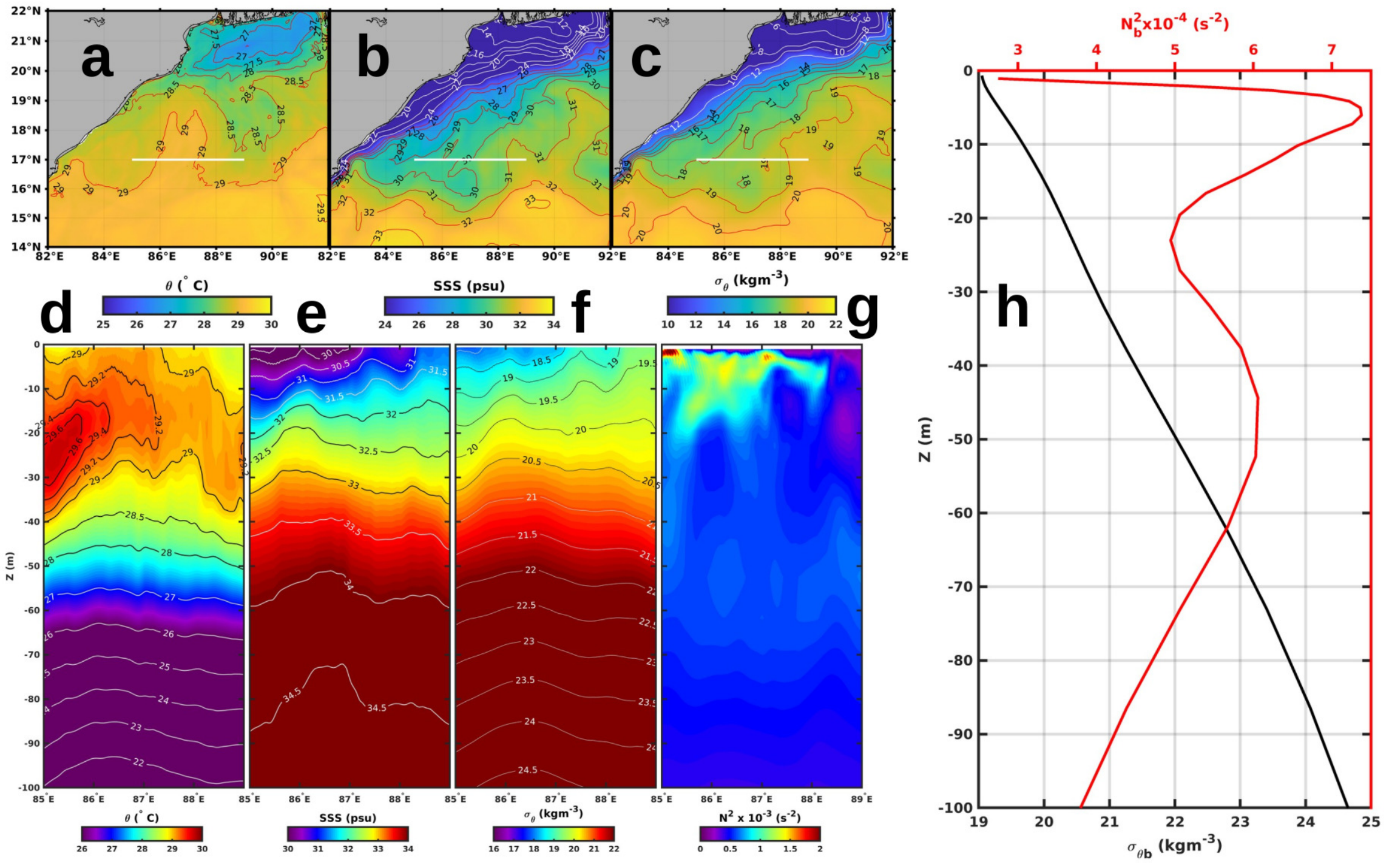}}
	\caption{(a), (b), (c), and (d) shows the mean potential temperature ($\theta$), sea-surface salinity (SSS), and the potential density ($\sigma_\theta$) averaged over October to December, at Z=-0.6 m, respectively. (d), (e), (f), and (g) shows the mean vertical profile of $\theta$, salinity, $\sigma_\theta$, and square of Brunt-v\"{a}is\"{a}l\"{a} frequency (N$^2$) averaged over the same period. (h) show the vertical profile of mean $\sigma_{\theta b}$ (black), N$_b^2$ (red)---time-mean for the whole season and area-mean over the box: $85^\circ$E-$89^\circ$E, $14^\circ$N-$18^\circ$N.}
	\label{s12}
\end{figure}

\begin{figure}[hbt!]
	\centerline{\includegraphics[width=\textwidth]{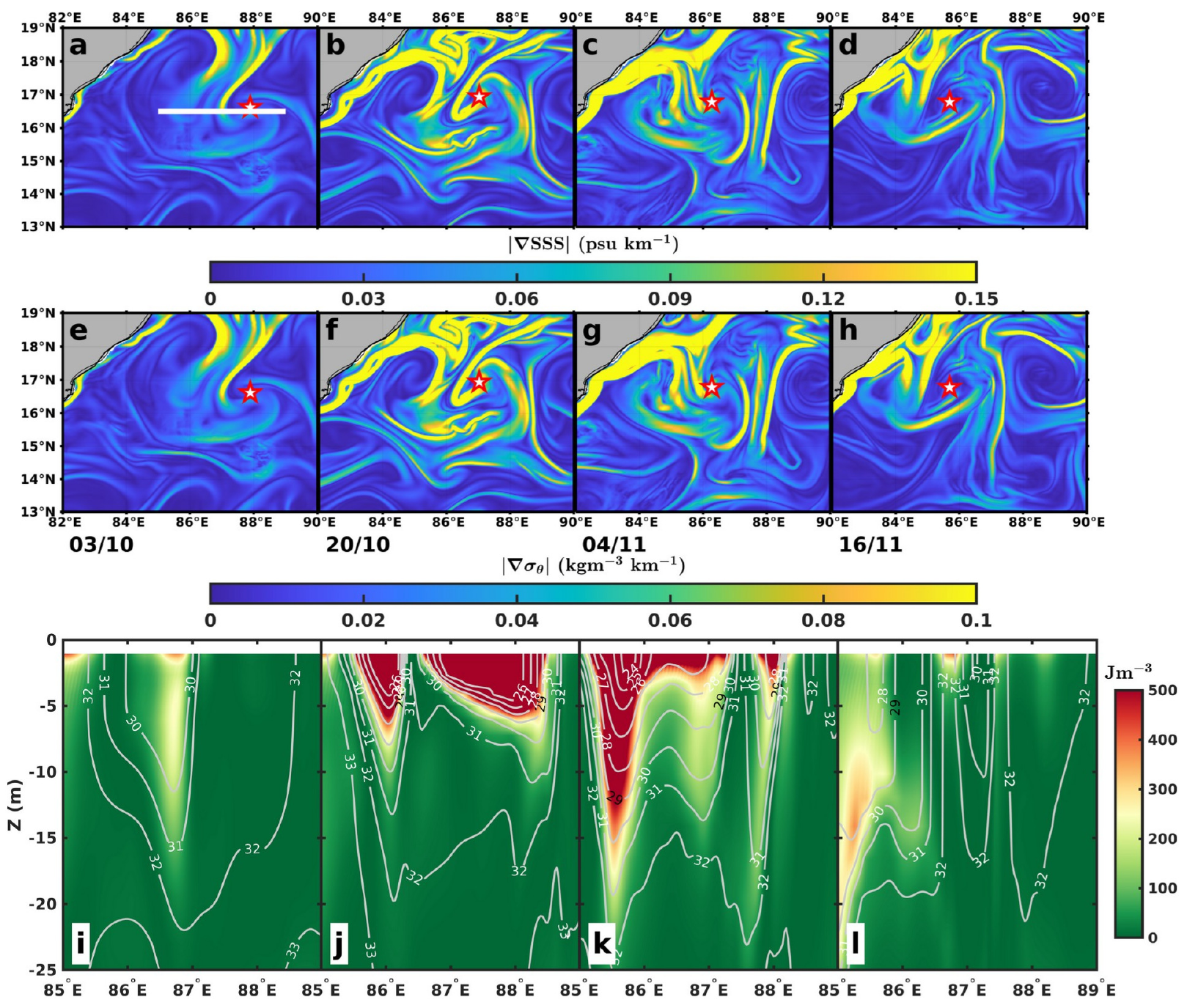}}
	\caption{(a)-(d), (e)-(h), and (i)-(l) represents the gradient of salinity, the gradient of potential density, and the vertical profile of the Eddy available potential energy (EPE) across the cross-section of the eddy. The star marks the center of the eddy based on the minima of the sea-level anomaly.}
	\label{s13}
\end{figure}

\begin{figure}[hbt!]
	\centerline{\includegraphics[width=\textwidth]{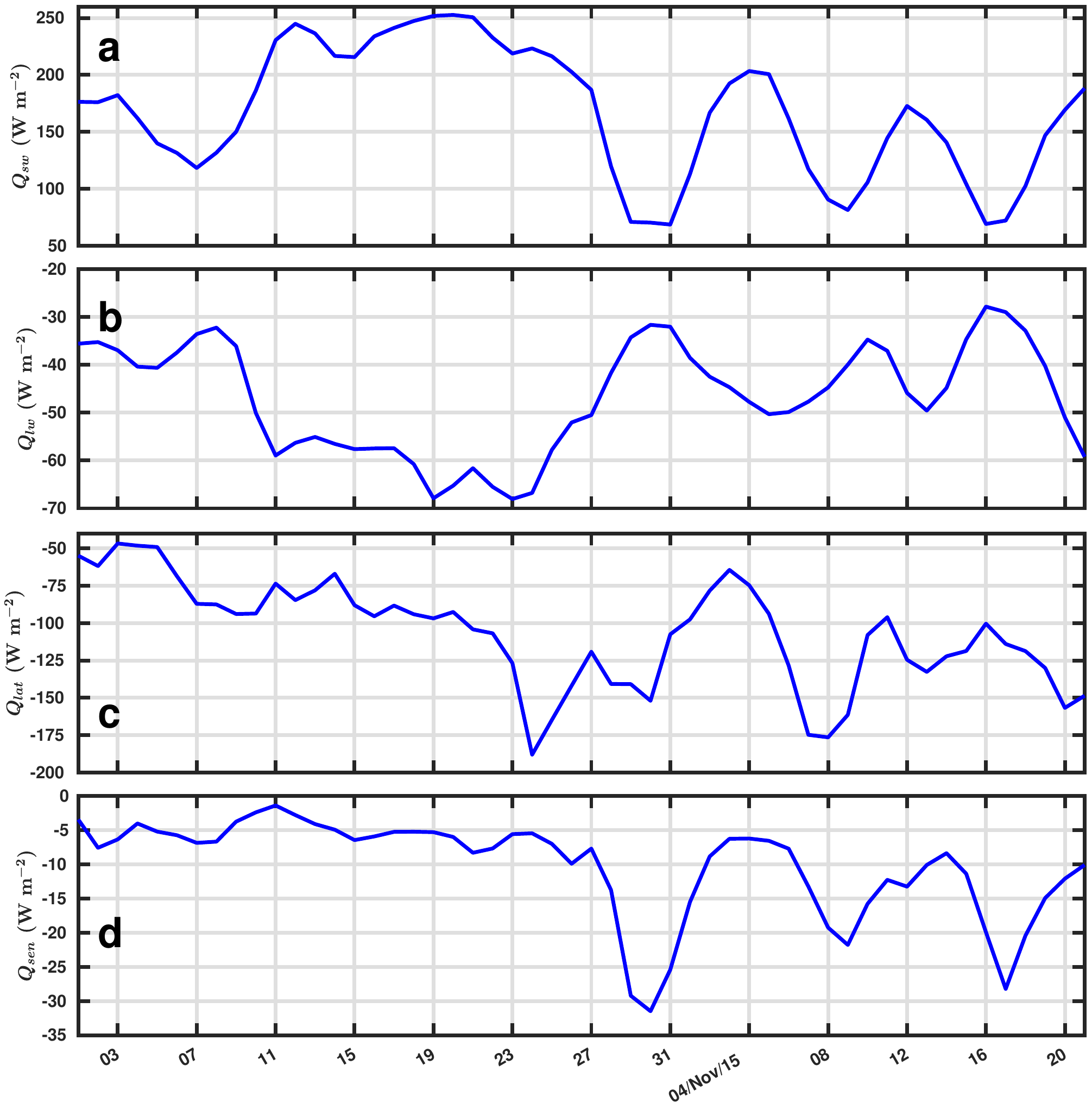}}
	\caption{(a), (b), (c), and (d) shows the net short-wave radiation ($Q_{sw}$), net long-wave radiation ($Q_{lw}$), latent heat flux ($Q_{lat}$), and sensible heat flux ($Q_{sen}$) averaged over the box: $85^\circ$E-$89^\circ$E, $14^\circ$N-$18^\circ$N shown from 20th October to 21st of November of 2015.}
	\label{s14}
\end{figure}

\begin{figure}[hbt!]
	\centerline{\includegraphics[width=\textwidth]{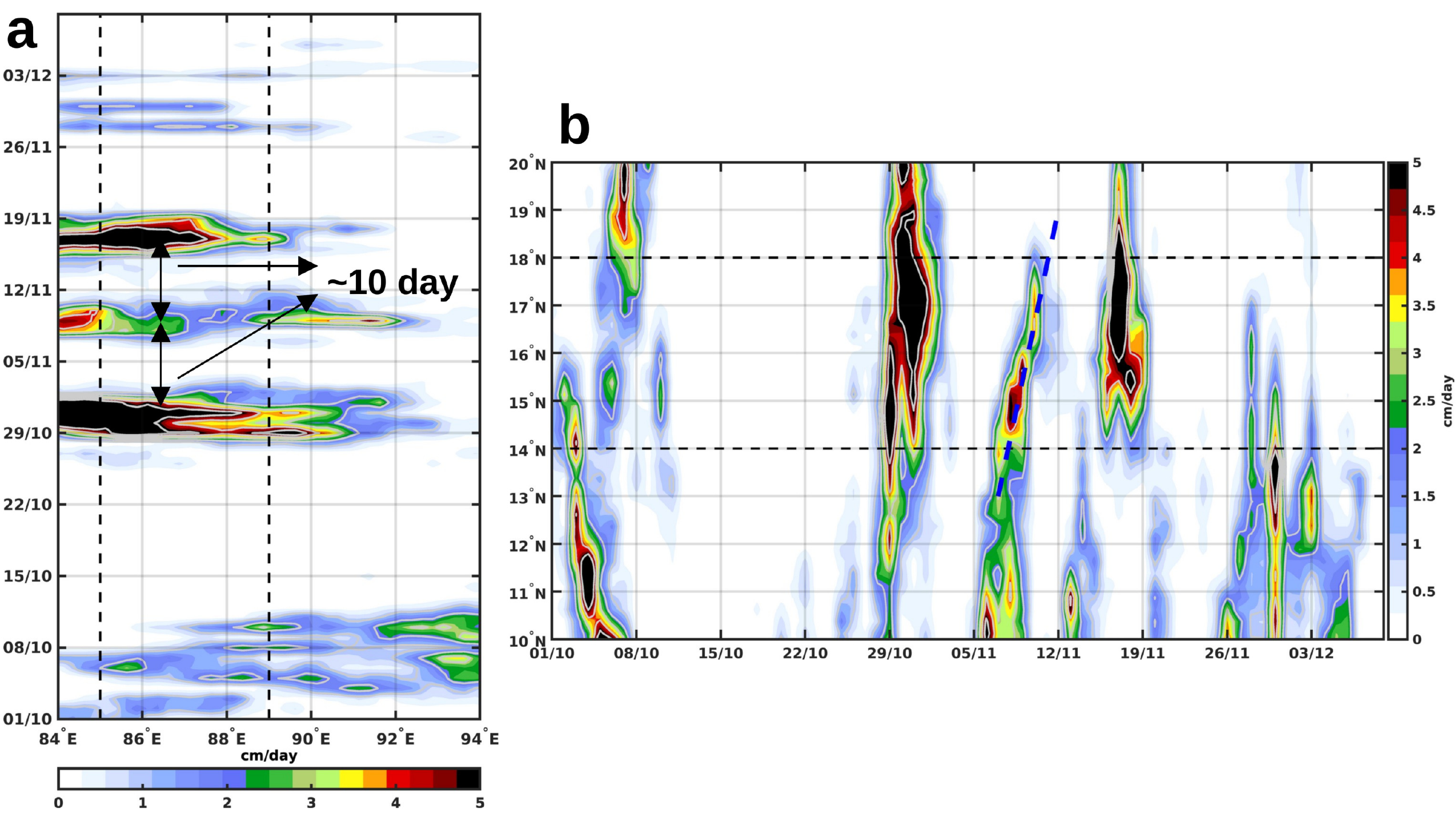}}
	\caption{(a) and (b) shows the Hovm\"{o}ller, latitude-time diagram of precipitation rate (cm day$^{-1}$) averaged over 14$^\circ$N-18$^\circ$N, 85$^\circ$E-89$^\circ$E, respectively from 1/10---9/12 of 2015.}
	\label{s15}
\end{figure}

\bibliography{references_nihar.bib}

\end{document}